\documentclass[sigconf]{acmart}

\usepackage{acmart-taps}
\usepackage{graphicx}
\usepackage{multirow}
\usepackage{listings}
\usepackage{xcolor}
\usepackage{float}
\usepackage{booktabs}     

\newcommand{\level}[1]{%
  \begingroup
  \def\dotfilled{\scalebox{2}{$\bullet$}}%
  \def\dotempty{\scalebox{2}{$\circ$}}%
  \ifcase#1
    \dotempty\ \dotempty\ \dotempty\ \dotempty\ \dotempty%
  \or
    \dotfilled\ \dotempty\ \dotempty\ \dotempty\ \dotempty%
  \or
    \dotfilled\ \dotfilled\ \dotempty\ \dotempty\ \dotempty%
  \or
    \dotfilled\ \dotfilled\ \dotfilled\ \dotempty\ \dotempty%
  \or
    \dotfilled\ \dotfilled\ \dotfilled\ \dotfilled\ \dotempty%
  \else
    \dotfilled\ \dotfilled\ \dotfilled\ \dotfilled\ \dotfilled%
  \fi
  \endgroup
}

\AtBeginDocument{%
  }

\copyrightyear{2026}
\acmYear{2026}
\setcopyright{cc}
\setcctype{by}
\acmConference[CHI '26]{Proceedings of the 2026 CHI Conference on Human Factors in Computing Systems}{April 13--17, 2026}{Barcelona, Spain}
\acmBooktitle{Proceedings of the 2026 CHI Conference on Human Factors in Computing Systems (CHI '26), April 13--17, 2026, Barcelona, Spain}
\acmPrice{}
\acmDOI{10.1145/3772318.3790653}
\acmISBN{979-8-4007-2278-3/2026/04}

\newcommand{\imgH}{1.3cm}

\usepackage{chi26-384}

\newcommand{\leftpad}{\rule{0pt}{2.6ex}} 
\newcommand{\lsep}{\vspace{0.35ex}}      

\begin{document}

\title{GenFaceUI: Meta-Design of Generative Personalized Facial Expression Interfaces for Intelligent Agents}

\author{Yate Ge}
\orcid{0009-0008-6617-6689}
\affiliation{%
  \institution{College of Design and Innovation, Tongji University}
  \city{Shanghai}
  \country{China}
}
\affiliation{%
  \institution{School of Design, Southern University of Science and Technology}
  \city{Shenzhen}
  \country{China}
}
\email{geyate@tongji.edu.cn}

\author{Lin Tian}
\orcid{0009-0004-8036-1750}
\affiliation{%
  \institution{College of Design and Innovation, Tongji University}
  \city{Shanghai}
  \country{China}}
\email{2433532@tongji.edu.cn}

\author{Yi Dai}
\orcid{}
\affiliation{%
  \institution{Shanghai Research Institute for intelligent Autonomous Systems, Tongji University}
  \city{Shanghai}
  \country{China}}
\email{2210999@tongji.edu.cn}

\author{Shuhan Pan}
\orcid{0009-0005-7136-7314}
\affiliation{%
  \institution{University of Washington}
  \city{Seattle}
  \state{Washington}  
  \country{USA}}
\email{amyshuhan2021@163.com}

\author{Yiwen Zhang}
\orcid{0000-0002-0892-0763}
\affiliation{%
  \institution{School of Art and Design, Wuhan University of Technology}
  \city{Wuhan} 
  \country{China}}
\email{zhangyw@whut.edu.cn}

\author{Qi Wang}
\orcid{0000-0002-2688-8306}
\affiliation{%
  \institution{College of Design and Innovation, Tongji University}
  \city{Shanghai}
  \country{China}
}
\email{qiwangdesign@tongji.edu.cn}

\author{Weiwei Guo}
\authornote{Xiaohua Sun and Weiwei Guo are the corresponding authors.}
\orcid{0000-0001-5037-0972}
\affiliation{%
  \institution{College of Design and Innovation, Tongji University}
  \city{Shanghai}
  \country{China}
}
\affiliation{%
  \institution{State Key Laboratory of General Artificial Intelligence, BIGAI}
  \city{Beijing}
  \country{China}
}
\email{weiweiguo@tongji.edu.cn}

\author{Xiaohua Sun}
\authornotemark[1]
\orcid{0000-0002-9206-628X}
\affiliation{%
  \institution{School of Design, Institute of Robotics Research, Southern University of Science and Technology}
  \city{Shenzhen}
  \country{China}
}
\email{sunxh@sustech.edu.cn}

\renewcommand{\shortauthors}{Ge et al.}

\begin{abstract}

This work investigates generative facial expression interfaces for intelligent agents from a meta-design perspective. We propose the Generative Personalized Facial Expression Interface (GPFEI) framework, which organizes rule-bounded spaces, character identity, and context--expression mapping to address challenges of control, coherence, and alignment in run-time facial expression generation. To operationalize this framework, we developed GenFaceUI, a proof-of-concept tool that enables designers to create templates, apply semantic tags, define rules, and iteratively test outcomes. We evaluated the tool through a qualitative study with twelve designers. The results show perceived gains in controllability and consistency, while revealing needs for structured visual mechanisms and lightweight explanations. These findings provide a conceptual framework, a proof-of-concept tool, and empirical insights that highlight both opportunities and challenges for advancing generative facial expression interfaces within a broader meta-design paradigm.

\end{abstract}



\maketitle

\section{Introduction}

\begin{figure*}
  \centering
  \includegraphics[width=1\textwidth]{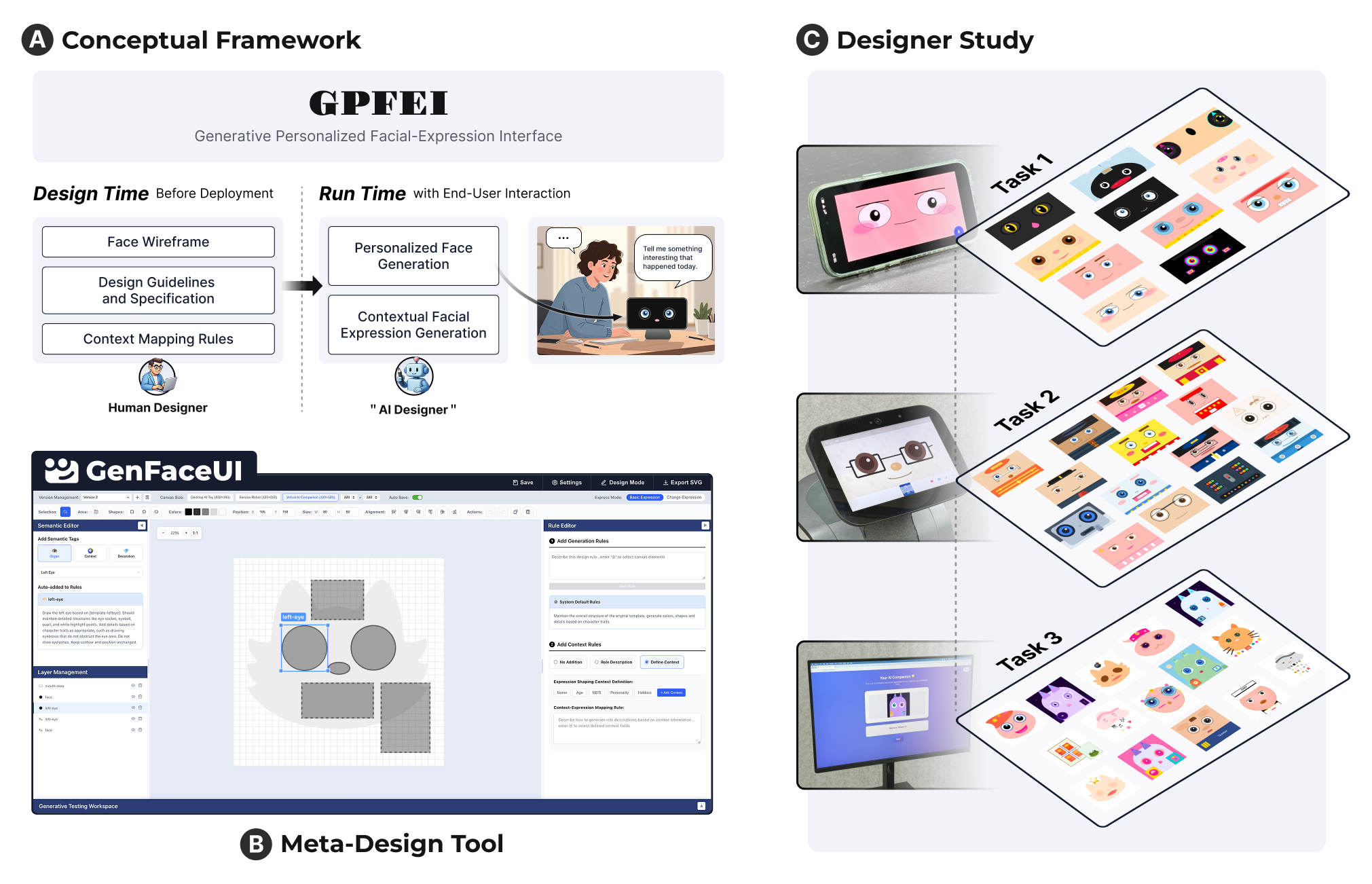}  
  \caption{Overview of the study: (A) the proposed GPFEI (Generative Personalized Facial Expression Interface) framework for intelligent agents; (B) \textsc{GenFaceUI}, a meta-design tool that operationalizes GPFEI; and (C) a qualitative study with designers using \textsc{GenFaceUI} across three design tasks.}
  \Description{The figure presents an overview of the research. Panel A illustrates the GPFEI (Generative Personalized Facial Expression Interface) framework, which structures generative personalized facial expression interfaces for intelligent agents through design-time specifications and run-time variation. Panel B shows \textsc{GenFaceUI}, a meta-design tool that implements this framework by enabling designers to author elements, semantic tags, and rules. Panel C depicts outputs from a qualitative study in which twelve designers engaged with three tasks---designing a basic chatbot face, customizing roles for a service robot, and creating a personalized AI companion---demonstrating how GPFEI can support diverse applications.}
  \label{fig:teaser}
\end{figure*}

Facial expressions are a key component of expressive intelligent agents.
In addition to signaling emotions in human-like forms, they improve usability, strengthen user engagement and trust, and support task performance \cite{el-nasr_emotionally_1999,torre_effect_2019}. They also shape perceptions of agency and characteristic traits, allowing users to recognize consistent roles and behaviors in the agent \cite{hara_use_1995,broadbent_robots_2013}. These interfaces have been applied across multiple domains, including social robots in physical environments \cite{breazeal_designing_2002,mishra_real-time_2023,antony_xpress_2025,serfaty_generative_2023,hong_development_2025} and virtual agents embedded in screen-based settings \cite{lee_irl_2025,leong_dittos_2024,rings_empathy_2024,jolibois_multimodal_2023,alam_text-based_2017}. In both contexts, screen-rendered facial expressions provide designers with broad flexibility to define character style and social cues, whether displayed on robot hardware or embodied as digital avatars, thereby enabling intelligent agents to maintain adaptive and coherent expressive identities \cite{kalegina_characterizing_2018}.

The rise of large language models (LLMs) has led to their increasing integration into intelligent agents to support expressive behaviors that are more flexible, personalized, and aligned with context \cite{wang_aint_2024,serfaty_generative_2023,mahadevan_generative_2024,ge_gencomui_2025}. 
Their enhanced context understanding, reasoning, and code generation capabilities further extend how interfaces can be produced and adapted at run time.
Interfaces generated at run time in response to interaction context have been discussed as an emerging user interface paradigm \cite{lee_towards_2025,chen_generative_2025,cao_generative_2025}, whose plasticity and adaptivity increase customizability and contextual adaptation \cite{lee_towards_2025}. For facial expression interfaces specifically, on-the-fly LLM-driven generation opens opportunities to improve context-sensitive affect display as well as functional interactivity. Recent systems demonstrate LLM-based inference of context and the generation of facial expression control parameters, enabling context-appropriate dynamic changes \cite{antony_xpress_2025,serfaty_generative_2023,mishra_real-time_2023}. 

However, the design space afforded by LLMs for facial expression interfaces remains underexplored. 
Trained on large-scale code and vector-graphics corpora, LLMs can parse and produce UI code\cite{lu_ui_2023} and vector graphics (e.g., SVG\footnote{\url{https://developer.mozilla.org/en-US/docs/Web/SVG}}) \cite{yang_omnisvg_2025,wu_chat2svg_2024}. 
This opens possibilities such as inserting contextual semantic visual elements into the face \cite{schombs_facevis_2024} and enabling user customization of the interface at run time \cite{lee_implementation_2023}, capabilities that may strengthen both expressivity and interaction quality when appropriately constrained.
At the same time, work on Generative UI (GenUI) reports challenges for real-world deployment, including issues of usability, consistency, and bias \cite{lee_towards_2025}.
As an illustrative case, the \textit{Xpress}\cite{antony_xpress_2025} study highlights the risk of inappropriate behaviors in sensitive contexts such as mental-health support or interactions with children, and recommends treating such systems as human-in-the-loop co-creation processes tailored to scenario requirements to mitigate these risks.

In sum, generative facial expression interfaces introduce a unique set of design challenges. Because screen-rendered expressions are generated at run time rather than fully specified at design time, designers must establish rules and constraints that bound and steer generation. They must also anticipate deployment behavior through model-in-the-loop, context-sensitive testing and inference. Addressing this gap requires dedicated tools and methods that support activities such as rule authoring, constraint management, and context--rule mapping, along with theoretical accounts that clarify how designers can structure, test, and refine rule-bounded generative spaces at run time \cite{lee_towards_2025}.

To better understand and address the design challenges of generative facial expression interfaces, we adopt the lens of meta-design, which emphasizes designing the conditions for design \cite{fischer_meta-design_2004}. In this view, meta-designers provide infrastructures, rules, and constraints that enable a system to evolve and continue being designed at run time. The core tenet that ``future uses and problems cannot be fully anticipated at design time'' directly resonates with the technical characteristics of generative facial expression interfaces. Within such interfaces, the human designer engages in meta-design to guide and constrain the ``AI designer'', which generates interface variations on the fly in response to situational context. Meta-design provides a foundation to discuss how generative facial expression interfaces redistribute design agency, differentiate design activities across design time and run time, and re-frame the role of designers in guiding the evolution of expressive systems.

Building on this lens, we introduce a conceptual framework for generative personalized facial expression interfaces (GPFEI), grounded in the affordances of LLM-based generative methods. The framework emphasizes design time meta-design activities where designers specify elements, layouts, colors, shapes, and rules that bound and guide run time generation. At run time, the ``AI designer'' produces role-specific customization and context-sensitive variation, extending agents' expressive capacities while remaining coherent with design intent (Figure~\ref{fig:teaser}-A).

To operationalize this framework, we developed \textsc{GenFaceUI}, a proof-of-concept meta-design tool that allows designers to compose templates, add semantic tags, define rules, and iteratively test outcomes (Figure~\ref{fig:teaser}-B). We used \textsc{GenFaceUI} as both an implementation and a research tool, examining through qualitative methods how designers engaged with meta-design, what challenges they faced, and what support is needed for practice. 
In this paper, we surface these challenges to frame generative facial expression interfaces as a distinct design problem, and use our findings to articulate implications for meta-design tools and for advancing generative facial expression interfaces within a broader meta-design paradigm (Figure~\ref{fig:teaser}-C).

In summary, this paper makes the following contributions:

\begin{itemize}
  \item We introduce the Generative Personalized Facial Expression Interface (GPFEI) framework for the meta-design of generative facial expression interfaces in intelligent agents, articulating how rule-bounded design spaces, character identity, and context-driven variation can be integrated.
  \item We contribute \textit{\textsc{GenFaceUI}}, a proof-of-concept tool that serves both as a generative facial expression interface design environment and as a probe for examining how designers engage in meta-design practices.
  \item We provide empirical insights from a qualitative study with designers (N=12), highlighting how they experience meta-design of generative interfaces, what challenges emerge, and what design implications follow for future tools and real-world deployments.
\end{itemize}

\section{Related Work}

\subsection{Expressive Agents and Facial Expression Interfaces}
The ability to express emotions is central to creating believable and engaging intelligent agents \cite{el-nasr_emotionally_1999}. Facial expression interfaces are therefore a key component: beyond conveying affect in a human-like manner, they enhance usability, foster engagement, build trust, and support task performance \cite{el-nasr_emotionally_1999,torre_effect_2019}. They have been widely applied in both social robots \cite{breazeal_designing_2002,mishra_real-time_2023,antony_xpress_2025,serfaty_generative_2023,hong_development_2025} and virtual agents \cite{lee_irl_2025,leong_dittos_2024,rings_empathy_2024,jolibois_multimodal_2023,alam_text-based_2017}.

Prior approaches to facial expression interfaces have been largely asset-authored and template-driven, where designers predefine fixed libraries or handcrafted rules before deployment. Such methods are labor-intensive and limited in expressive coverage, particularly for subtle or mixed emotional states \cite{herdel_drone_2021}. Subsequent machine-learning approaches automated expression generation through dimensional emotion models \cite{trovato_development_2012}, but remained tied to specific datasets and lacked transferability across platforms, restricting adaptability and expressiveness in real use. More recent work integrates LLMs to infer context and generate low-level expression parameters for context-appropriate variation \cite{antony_xpress_2025,serfaty_generative_2023,mishra_real-time_2023}. A notable example is the \textit{Xpress} system \cite{antony_xpress_2025}, which demonstrates the feasibility of mapping natural language understanding to facial-parameter control in screen-rendered, vector-based expression interfaces. Yet, such solutions still focus narrowly on parameter modulation, without leveraging the broader design opportunities of LLMs---such as embedding contextual visual cues \cite{schombs_facevis_2024} or enabling user-driven customization at run time \cite{lee_implementation_2023}.

In summary, existing work has established the value of facial expression interfaces and explored LLM-driven expression control, yet the broader design space of generative facial expression interfaces remains underexplored. Our work addresses this gap by investigating new interface design possibilities through a design tool and an empirical study of designers' meta-design practices.

\subsection{LLM-Based Expressive Behavior and Generative Interfaces}

Recent advances in LLMs have expanded their role beyond text generation to multimodal expressive behaviors. These systems can transform language inputs into context-aware expressive actions \cite{mahadevan_generative_2024,huang_emotion_2024}, facial expressions \cite{wang_aint_2024,kim_contextface_2025}, and visual representations \cite{ge_gencomui_2025,sonawani_sisco_2024}, enabling agents to align communicative form with content.
Such expressive behaviors are increasingly integrated into various forms of interactive intelligent agents, including social robots, digital humans, avatars, virtual companions, and game non-player characters (NPCs) \cite{allgeuer_when_2024,markelius_empirical_2024,brito_integrating_2025,do_implementing_2025,lee_irl_2025,galatolo_simultaneous_2025,chen_emo-avatar_2025,lai_llm-driven_2025,ozkaya_how_2025}.
These developments enhance agents' expressive capabilities, enabling them not only to determine \textit{what} to communicate but also \textit{how} to convey and enact it across multiple expressive modalities.

Concurrently, HCI research has begun to conceptualize generative user interfaces as a new interaction paradigm. Unlike conventional interfaces specified at design time, generative user interfaces are generated on the fly at run time in response to user input and contextual factors \cite{chen_generative_2025,lee_towards_2025,leviathan_generative_nodate}. This paradigm emphasizes dynamic responsiveness and personalization, integrating design-time definitions with run-time synthesis to support more adaptive and user-aligned experiences.

Together, these strands LLM-driven expressive behaviors and the emerging paradigm of generative user interfaces suggest new opportunities for intelligent agents: generative expressive behaviors driven by LLMs can be coupled with generative interfaces to produce coherent, adaptive facial expression--mediated interactions. Generative facial expression interfaces occupy a special position at this intersection, functioning both as a graphical user interface and as an expressive modality for intelligent agents. While prior work has demonstrated technical feasibility, the broader design space of such interfaces, considering key aspects such as their design frameworks, mechanisms for controllability, and implications for designer agency, remains underexplored. Our work addresses this gap by examining generative facial expression interfaces as a new design paradigm.

\begin{figure*}
    \centering
    \includegraphics[width=1\linewidth]{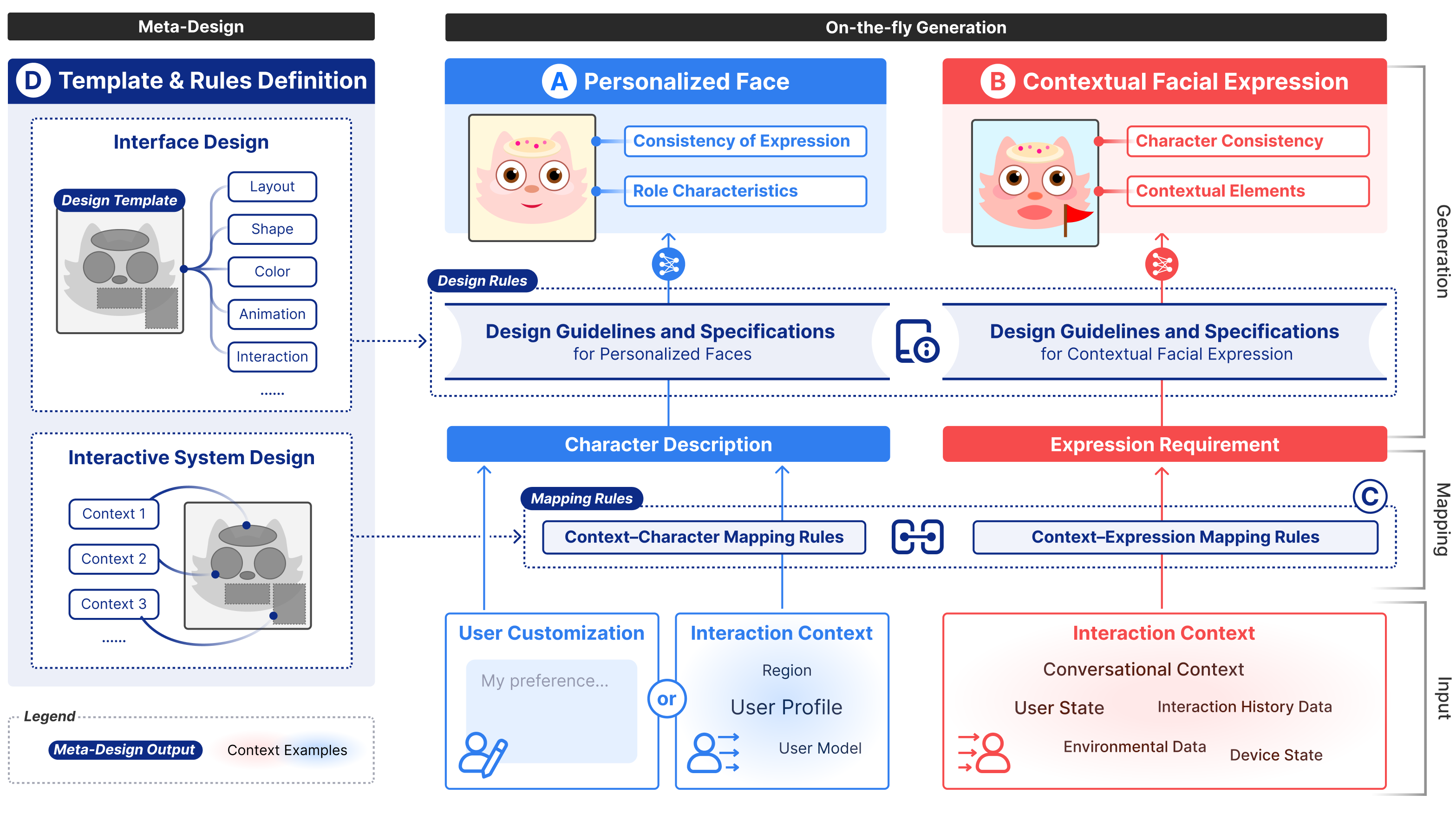}
    \caption{The GPFEI framework, highlighting personalized face customization and context-driven facial expressions under meta-design.}
    \Description{A two-lane block diagram. The top lane (design time) contains boxes for SVG Template, Semantic Tags, Constraints, and Context--Rule Mapping with arrows pointing to the bottom lane (run time). The bottom lane has two branches: one synthesizes a personalized base face from character/profile inputs; the other maps contextual inputs (e.g., dialogue, system state) to expression variations, yielding coherent facial outputs.}
    \label{fig:conceptualFramework}
\end{figure*}

\subsection{Meta-design of Generative Interface }

Recent advances in AI-assisted UI and UX design have integrated large language models into workflows across ideation, prototyping, evaluation, and UI code generation \cite{muehlhaus_interaction_2024,lu_ui_2023,yuan_maxprototyper_2024,duan_towards_2023,jing_layout_2023,duan_generating_2024,chen_designcoder_2025,ahmed_role_2025}. Yet these tools largely generate fixed, design-time interface artifacts rather than interfaces that adapt at run time. Even systems that introduce AI behavior simulation, such as PromptInfuser and Canvil \cite{petridis_promptinfuser_2023,petridis_promptinfuser_2024,feng_canvil_2025}, still operate at the prompt or functional level and do not treat run-time UI variation as an expressive behavior of the AI system. In parallel, work on generative user interfaces conceptualizes GenUI as a co-creative paradigm spanning design-time configuration and run-time adaptation \cite{lee_towards_2025}, but its implications for design practice remain underexplored.

A central challenge of generative interfaces lies in the redistribution of design responsibilities between human designers and AI systems. Instead of producing fixed assets ahead of time, designers must delimit the generative space of LLMs while preserving flexibility for creative variation at run time. This shift foregrounds questions of agency and concerns the balance between design intent, system autonomy, and end-user contributions in shaping interaction outcomes. Prior work has highlighted issues such as representation fidelity, temporal context, and adaptivity or plasticity \cite{lee_towards_2025}, yet empirical studies examining how these dynamics reshape design practice remain scarce. Although some LLM-driven UI tools embed rules, grammars, or constraints to guide generation \cite{lu_ui_2023,jing_layout_2023,chen_designcoder_2025}, they seldom support designers in explicitly shaping or refining the underlying generative space---especially when interfaces evolve at run time.

Meta-design provides a useful theoretical foundation for understanding and addressing these challenges. Fischer's meta-design paradigm emphasizes designing the conditions for design and envisions systems that evolve through ongoing co-evolution between users and AI systems \cite{fischer_meta-design_2004}. More recent work extends this perspective by highlighting the need for methods that empower end-users through AI-supported development and enable continuous adaptation of interactive systems \cite{barricelli_advancing_2024}. 
Viewed through this lens, generative interfaces require tools that help designers configure and evolve the expressive space of LLM-driven systems, including the behaviors that unfold at run time. This need extends beyond current AI-assisted design tools, which remain focused on design-time functionality.

In summary, while prior research has conceptualized generative interfaces and identified their technical and design concerns, 
and while many AI-assisted design tools now support different phases of design workflows, little empirical work examines how meta-design can shape the redistribution of design agency or help designers define the generative spaces that govern run-time interface behavior.
Our study contributes by investigating generative facial expression interfaces as a site for exploring these developments and provides empirical insights into how designers engage with meta-design in the context of this emerging interaction paradigm and how they iteratively negotiate and refine rules and constraints to align LLM-driven expressive behaviors with design intent at run time.

\section{GPFEI: Conceptual Framework}
\label{sec:GPFEI}

We present the theoretical rationale underlying the Generative Personalized Facial Expression Interface (GPFEI) framework, describe its core components and relationships, and outline design challenges that inform the development of a meta-design tool for the GPFEI (Figure~\ref{fig:conceptualFramework}).

\subsection{Motivation of the Framework}

The GPFEI framework aims to clarify how facial expressions for intelligent agents can be generated dynamically at run time while remaining bounded by design-time specifications. Its motivation is to support both character customization and contextual expressivity, and to position meta-design as the mechanism that constrains and steers generative outcomes. In doing so, the framework serves as an exploratory lens to probe the design space of generative facial expression interfaces and the emerging paradigms of designer practice, offering initial guidance rather than a comprehensive account.

\subsection{Theoretical Rationale}

The theoretical rationale of the GPFEI framework builds on Fischer's articulation of adaptive systems (grounded in AI) and adaptable systems (grounded in end-user development \cite{lieberman_end-user_2006}) not as mutually exclusive    categories, but as complementary components in a symbiotic relationship \cite{fischer_adaptive_2023}. Adaptive systems autonomously adjust their behavior through context-aware models of users and tasks, whereas adaptable systems empower users to modify, extend, and steer functionality to meet emergent needs. 
Fischer argues that the future of socio-technical environments lies in integrating automation and human agency into a co-evolutionary process.
We draw on this rationale to position generative personalized facial expression interfaces as a domain where adaptive and adaptable elements must be carefully balanced, enabling both dynamic expressivity at run time and designer-driven control at design time.

In the context of facial expression interfaces for intelligent agents, this dual perspective provides a conceptual grounding. Personalization through character customization corresponds to adaptable systems: users or designers deliberately configure traits, visual styles, and expressive options to strengthen identity connections and foster engagement and trust \cite{gao_learning_2022,oliva_design_2025,lee_implementation_2023}. Context-sensitive expression corresponds to adaptive systems: facial interfaces dynamically alter expressions or integrate contextual elements (e.g., emojis, status indicators) in response to dialogue history, system state, or environmental cues \cite{wang_design_2022,ebisu_see_2025,ben_youssef_towards_2015}. Taken together, these two dimensions enable situated expressivity that transcends static emotion categories and supports more contextualized interaction.

This rationale resonates with Fischer's view of meta-design as ``designing the conditions for design,'' acknowledging that future uses cannot be fully anticipated at design time \cite{fischer_meta-design_2004}. Extending this perspective, recent work highlights the role of AI in empowering end-user development and enabling co-evolutionary, human-centered systems \cite{barricelli_advancing_2024}. 
Within GPFEI, human designers define guiding constraints that shape the generative space, while the AI designer produces situated variations at run time. 
This grounding situates the framework in a human--AI co-design perspective, where run-time personalization by the AI is guided and bounded by the design-time rules and constraints specified by human designers.

\subsection{Generative Personalized Facial Expression Interface (GPFEI)}
As shown in Figure~\ref{fig:conceptualFramework}, GPFEI is an interaction framework that specifies how screen-rendered facial expressions are generated at run time under design-time specifications. It comprises four components:

\textbf{(1) Personalized Face Generation (Figure~\ref{fig:conceptualFramework}-A).} 
At design time, designers define a design template (layout, shapes, palette, animation, interaction style) and a rule set that bounds admissible variations and encodes style invariants. At run time, a \emph{character description}---from user customization and/or profile/region/deployment context---is mapped to template parameters via context--character mapping rules, and the face is instantiated within the allowed ranges. The output is the personalized, character-specific base face determined by the input description and guaranteed to satisfy the design guidelines and constraints.

\textbf{(2) Contextual Facial Expression (Figure~\ref{fig:conceptualFramework}-B).} 
Based on the personalized base face, expression variation is generated at run time according to rules and constraints predefined by designers. Interaction context---including dialogue history, user state, environmental data, and device status---is translated into an \emph{expression requirement} and mapped through context--expression rules to parameter updates on the base face. The output is a context-aligned expression that adapts dynamically to situational factors while remaining consistent with the predefined design guidelines and the character's role identity.

\textbf{(3) Context Mapping (Figure~\ref{fig:conceptualFramework}-C).} 
To connect interaction inputs with generative behaviors, designers define context--rule mappings at design time. These mappings specify how different contextual factors (e.g., dialogue intent, system state, environment signals, or user profile attributes) are transformed into parameters for face personalization or expressive variation. At run time, incoming context is parsed through these predefined mappings to determine which rules apply and how the base face should be updated. In this way, context mapping serves as the mechanism that operationalizes the link between situational inputs and consistent generative outputs.

\textbf{(4) Meta-design (Figure~\ref{fig:conceptualFramework}-D).} 
Meta-design refers to the design-time activity through which human designers define the generative space itself. Rather than producing fixed outputs, designers author templates, constraints, and context--rule mappings that delimit how (1) personalized faces and (2) contextual expressions can be generated, and how (3) context inputs are linked to parameter changes. These specifications establish the structural boundaries, stylistic invariants, and transformation rules that govern run-time behavior. In this way, meta-design provides the foundation that ensures all generative outputs remain adaptive and personalized while aligned with design intent\cite{karwowski_design_2019}.

Together, these four components outline how design-time specifications guide run-time generation, linking personalized face construction, contextual variation, and mapping mechanisms under the overarching activity of meta-design.

\subsection{Design Challenges for Meta-Design Tool}
Building on the GPFEI framework, we identified several key design challenges for developing a meta-design tool that supports designers in practice:

\begin{itemize}
  \item \textbf{DC1: Visual Editing Workspace.} Designers are accustomed to tools like Figma or Illustrator; the tool should provide a familiar visual canvas for directly editing and arranging elements. 
  \item \textbf{DC2: Structured Rule Authoring.} Rules and design intent must be specified through structured, interpretable interfaces rather than complex prompt engineering. 
  \item \textbf{DC3: Preserving Designer Agency.} The tool should ensure that designers' constraints and creative intent continue to guide generative outcomes over time. 
  \item \textbf{DC4: Run-time Simulation and Feedback.} Designers need mechanisms to simulate contextual inputs, observe generative results, and receive effective feedback to iteratively refine their specifications. 
\end{itemize}

\begin{figure*}
    \centering
    \includegraphics[width=1\linewidth]{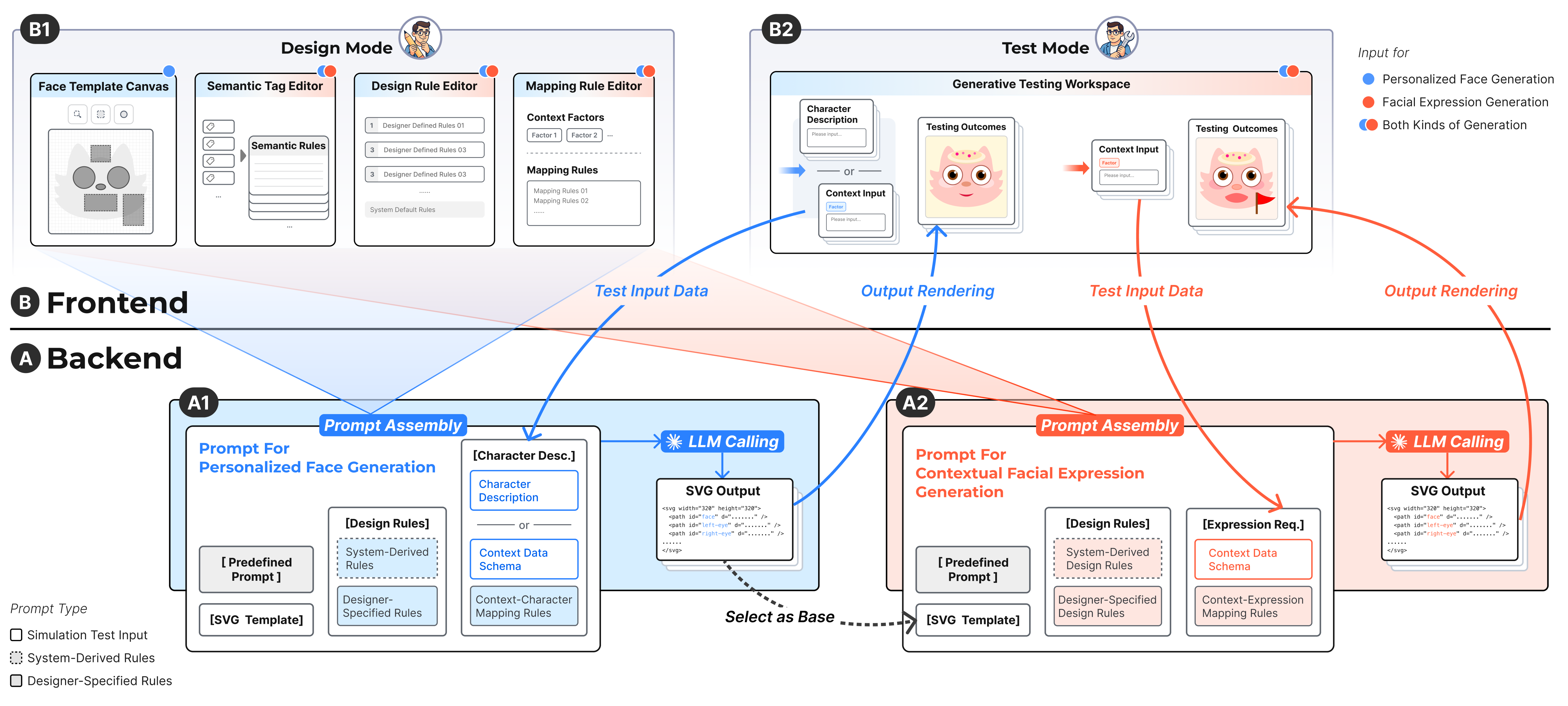}
    \caption{\textsc{GenFaceUI} Architecture aligned with GPFEI's two meta-design capabilities: personalized face generation and contextual facial expression generation. The backend assembles prompts from four modules (Default Prompt, SVG Template, Design Rules, Context Rules), while the frontend provides Design Mode and Test Mode for iterative design--test--select.}
    \Description{A diagram with two capability lanes: personalized face generation outputs a selected base face that flows into contextual facial expression generation as the SVG Template seed. Prompt assembly modules appear on the backend, and the frontend exposes Design Mode for authoring and Test Mode for simulating inputs and inspecting outputs.}
    \label{fig:architecture}
\end{figure*}

\section{\textsc{GenFaceUI}: A Meta-Design Tool}

Taking the GPFEI framework and design challenges into comprehensive account, we present \textsc{GenFaceUI}, a \textit{meta-design tool} for generative facial expression interfaces. The system architecture is organized around two meta-design capabilities (Figure~\ref{fig:architecture}): \textit{personalized face generation} and \textit{contextual facial expression generation}, and the interface overview illustrates Design/Test Modes (Figure~\ref{fig:overview}). The remainder of this section summarizes the UI components and a brief workflow.

\subsection{System Architecture and Data Flow}

As Shown in Figure \ref{fig:architecture}, \textsc{GenFaceUI} organizes meta-design around \textbf{two capabilities} of the GPFEI framework: \textit{personalized face generation} and \textit{contextual facial expression generation} (Section~\ref{sec:GPFEI}). Designers first specify templates and rules for personalized face generation, then  \textit{select a base face} and test. The selected base face is injected directly into the contextual facial expression generation prompt as the seed [SVG Template], after which designers author context rules and test contextual outputs. This establishes a clear handoff: \textit{design $\to$ test (personalized) $\to$ select } $\to$ \textit{prompt injection $\to$ design $\to$ test (contextual)}.

At the technical level, the implementation mirrors this handoff: the frontend (React) provides the Face Template Canvas, Semantic Tag Editor, Design Rule Editor, Context Mapping Rule Editor in Design Mode (Figure \ref{fig:architecture}-B1), and the Runtime Expression Simulator in Test Mode (Figure \ref{fig:architecture}-B2); the backend (Node.js) persists projects and assembles phase-specific prompts (Figure \ref{fig:architecture}-A1, A2) from four modules---[Predefined Prompt], [SVG Template], [Design Rules], and [Context Rules]---for the generative backbone , powered by \texttt{claude-opus-4-20250514}(temperature fixed at~0 for consistency). Designs run in-browser and can be validated on local devices over the same network for realistic testing.

\subsubsection{\textbf{Template, Rules, and Prompt Assembly.}}
The backend (Figure \ref{fig:architecture}-A) assembles prompts from four modules: [Predefined Prompt], [SVG Template], [Design Rules], and [Context Rules]. This assembly acts as a \textit{translation} from structured meta-design specifications to executable prompt text. For personalized face generation, [SVG Template] originates from the Face Template Canvas and [Design Rules] constrain identity, palette, geometry, and style invariants. For contextual facial expression generation, [SVG Template] is the \textit{selected base face} and [Context Rules] map contextual factors to parameter updates. Rule precedence is resolved as: element-targeted rules \(>\) global rules \(>\) default rules. See Appendix~\ref{app:prompt} for full prompt templates.

\subsubsection{\textbf{Simulated Inputs and Output Inspection.}}
Designers use Test Mode (Figure \ref{fig:architecture}-B2) to simulate inputs and inspect outputs for both capabilities. For personalized face generation, they provide character descriptions or context inputs, review multiple outputs, and click \texttt{Select as Base} to fix the baseline. For contextual facial expression generation, they specify context factors (presets or custom), apply context--expression mapping rules, and observe generated facial expressions. Outputs can be compared, re-tested with varied inputs, and, if needed, trigger revisions to templates or rules before re-testing. Details of meta-design tool components appear in Section~\ref{sec:components}.

\subsection{Meta-Design Tool Components} 
\label{sec:components}
\begin{figure*}
    \centering
    \includegraphics[width=1\linewidth]{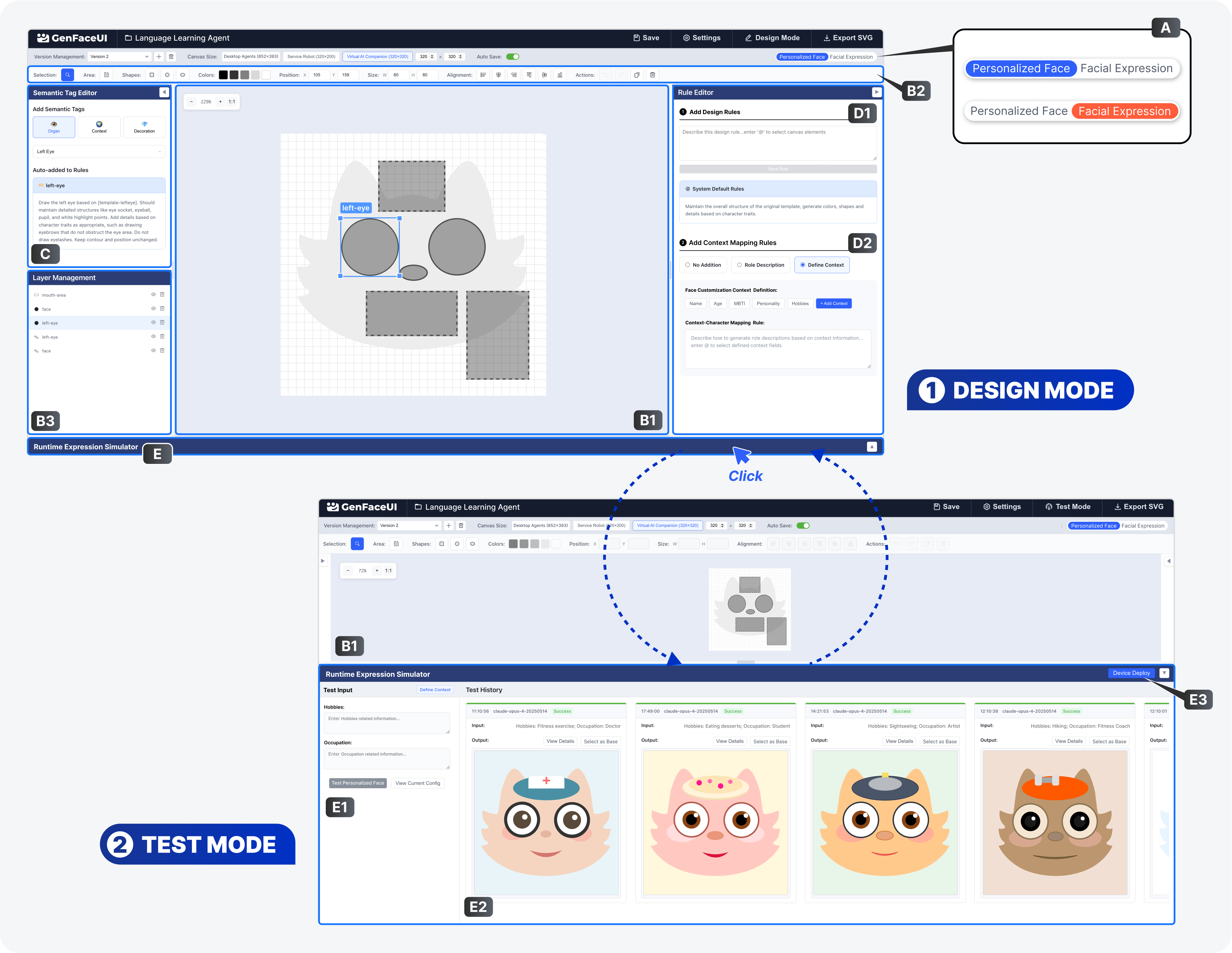}
    \caption{Interface overview with two modes: Design (1) and Test (2). In Design Mode, the Face Template Canvas (B1) and Toolbar (B2) compose the SVG template; the Semantic Tag Editor (C) applies tags and auto-adds default rules; the Design Rule Editor (D1) and Context Mapping Rule Editor (D2) define design and context-mapping rules. In Test Mode, designers provide context inputs (E1) and inspect generated outputs (E2) in the Runtime Expression Simulator (E), with optional device deploy (E3). The Phase Toggle (A) switches between \textit{personalized face generation} and \textit{contextual facial expression generation}.}
    \Description{A UI overview of GenFaceUI showing two panes: Design Mode with canvas, tag and rule editors, and Test Mode with a simulator for inputting context and viewing generated facial expressions.}
    \label{fig:overview}
\end{figure*}

The interface of \textsc{GenFaceUI} supports meta-design across two capabilities (Figure~\ref{fig:overview}-A)---\textit{personalized face generation} and \textit{contextual facial expression generation}---through two modes and five main components. The Design Mode (Figure~\ref{fig:overview}-1) and Test Mode (Figure~\ref{fig:overview}-2) allow switching between authoring and validation, while the components map directly to prompt modules: \textbf{Face Template Canvas} (Figure~\ref{fig:overview}-B1) $\rightarrow$ [SVG Template], \textbf{Semantic Tag Editor} (Figure~\ref{fig:overview}-C) $\rightarrow$ system-derived default element rules for [Design Rules], \textbf{Design Rule Editor} (Figure~\ref{fig:overview}-D1) $\rightarrow$ designer-specified rules for [Design Rules], \textbf{Context Mapping Rule Editor} (Figure~\ref{fig:overview}-D2) $\rightarrow$ [Context Rules], and \textbf{Runtime Expression Simulator}(Figure~\ref{fig:overview}-E) $\rightarrow$ simulated inputs and output inspection. Designers can quickly iterate by expanding or collapsing the Runtime Expression Simulator header.

\subsubsection{\textbf{Face Template Canvas (Figure~\ref{fig:overview}-B1).}} 
\begin{figure}
    \centering
    \includegraphics[width=\linewidth]{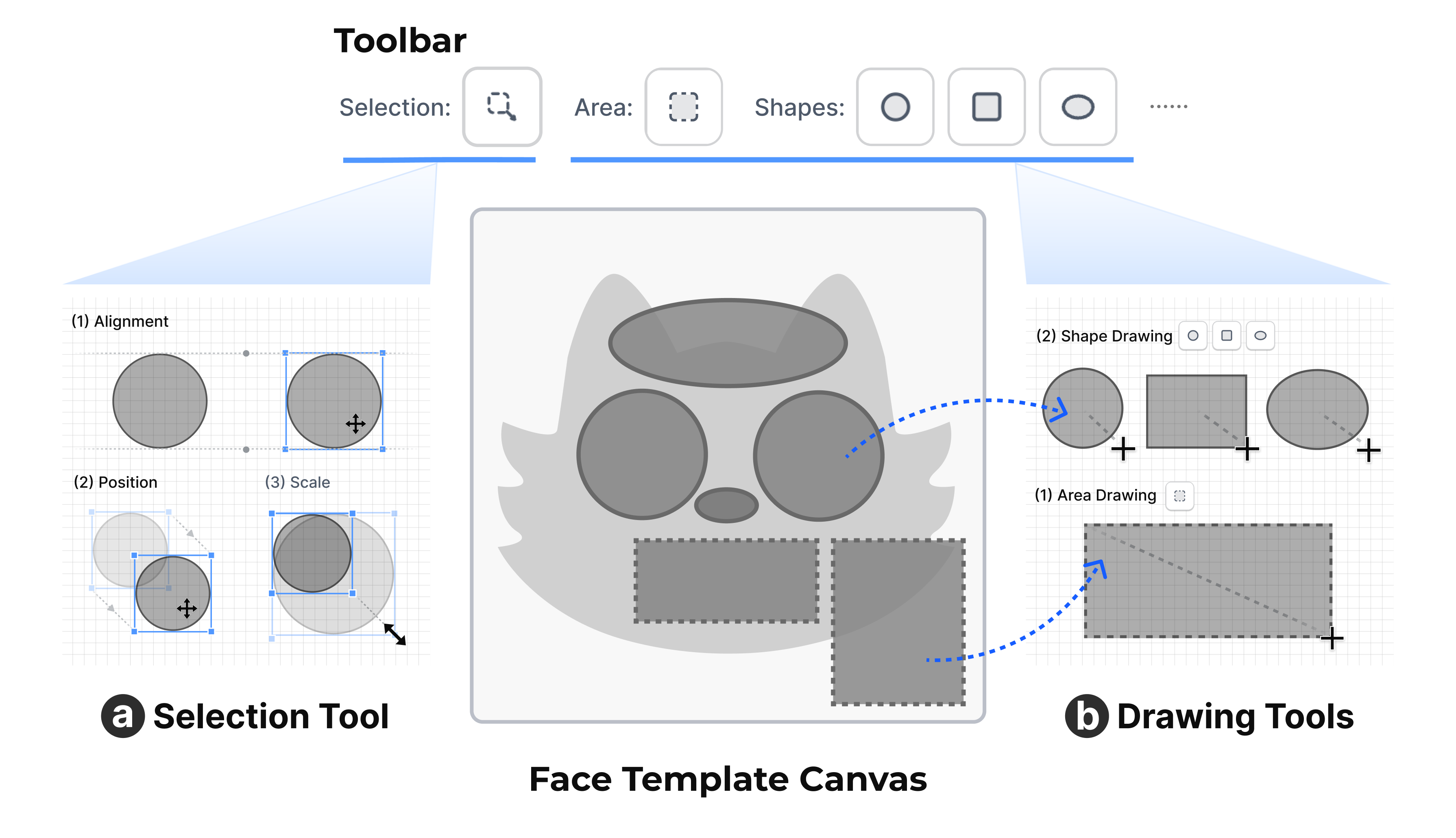}
    \caption{Face Template Canvas and Toolbar: (a) Selection Tool allows alignment, positioning, and scaling of elements; (b) Drawing Tools provide Area Drawing to define placement ranges and Shape Drawing to create geometric shapes.}
    \Description{Screenshot of the canvas for sketching the face template with selection and drawing tools visible.}
    \label{fig:canvas}
\end{figure}

The Face Template Canvas acts as the primary workspace where designers create prototype shapes (\textbf{DC1}) and define spatial constraints for expression elements. Two main drawing tools (Figure \ref{fig:canvas}-b) are provided in the Toolbar (Figure~\ref{fig:overview}-B2): \textbf{(1) Area Drawing}, which constrains the placement range of elements without fixing their specific shapes; and \textbf{(2) Shape Drawing}, which applies geometric primitives (circle, ellipse, rectangle) to strictly constrain element appearance. Designers can also import SVG contours from tools like Figma for more complex or distinctive forms. In addition, the Toolbar integrates other essential functions such as a selection tool for adjusting position, scale, and alignment of elements (Figure \ref{fig:canvas}-a), and a grayscale fill palette for clear visual distinction. The Element Management Panel (Figure~\ref{fig:overview}-B3) further supports reordering, hiding, and deleting elements. This component produces the \textbf{[SVG Template]} for prompt assembly.

\subsubsection{\textbf{Semantic Tag Editor (Figure~\ref{fig:overview}-C)}} 
\begin{figure}
    \centering
\includegraphics[width=\linewidth]{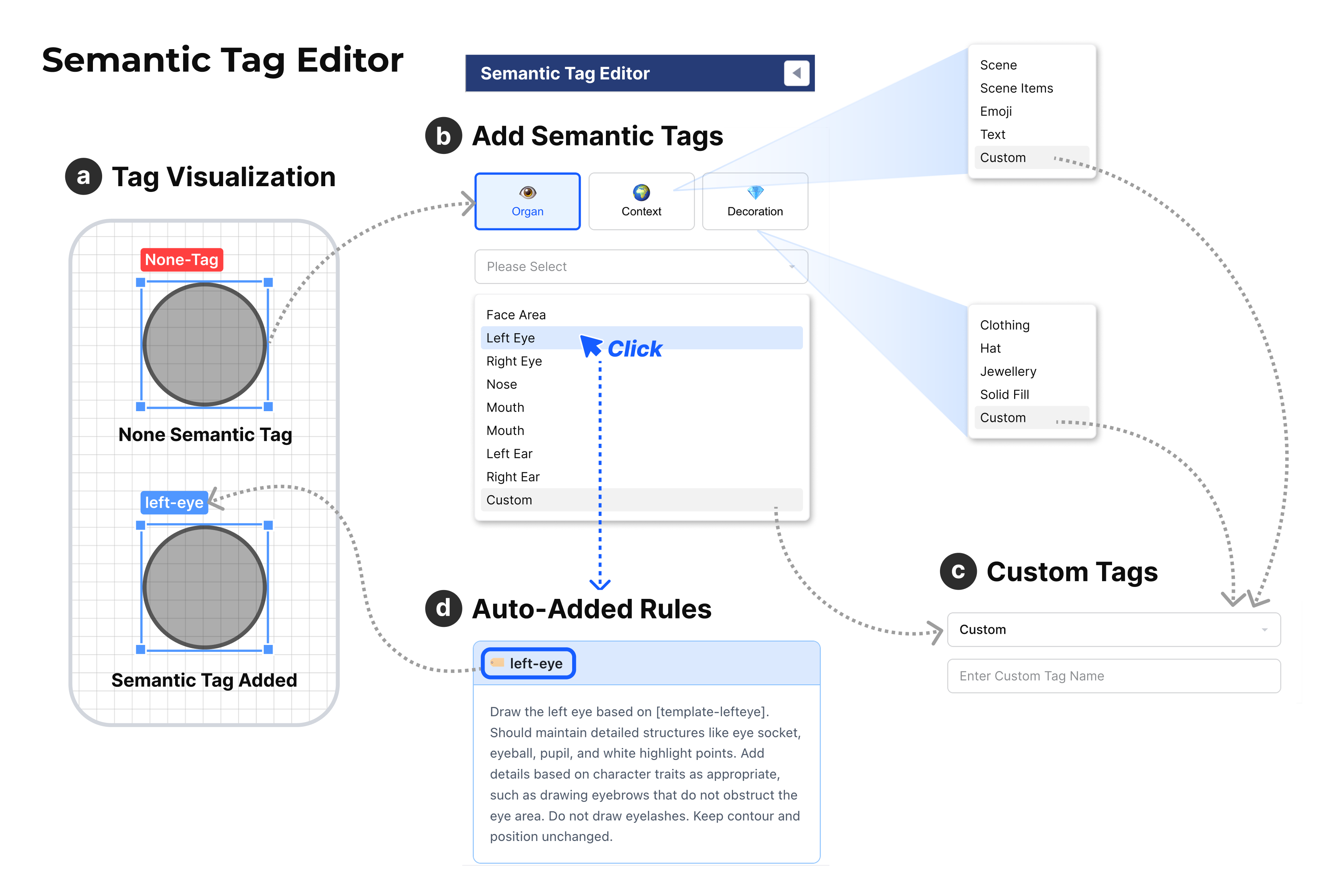}
    \caption{Semantic Tag Editor: users select an element (a), add semantic tags (b) from organ, decoration, or context categories or custom names (c), and review auto-added default rules (d) that constrain the element.}
    \Description{Screenshot of the semantic tagging interface with element selection, tag categories, and default rules panel.}
    \label{fig:semantic}
\end{figure}

The Semantic Tag Editor enables designers to add semantic tags to selected elements (Figure \ref{fig:semantic}-a) on the canvas. Once a tag is applied, the system automatically adds a default rule (Figure \ref{fig:semantic}-d, \textbf{DC2}) that specifies what the element is, its basic attributes, and spatial constraints that guide subsequent generative outputs (see in Appendix~\ref{app:semantic}). The Editor offers three categories of preset tags (Figure \ref{fig:semantic}-b) for quick selection: organ tags (e.g., face, eyes, nose) for core facial elements; decoration tags (e.g., clothing, hats, accessories) that enrich expressive variety; and context tags (e.g., scene, emoji, text) that allow elements to adjust dynamically to provided contexts. Designers can further create custom tags (Figure \ref{fig:semantic}-c, \textbf{DC3}) to support their creative needs. These defaults populate \textbf{[Design Rules]}.

\subsubsection{\textbf{Design Rule Editor (Figure~\ref{fig:overview}-D1).}} 
\begin{figure}
    \centering
\includegraphics[width=\linewidth]{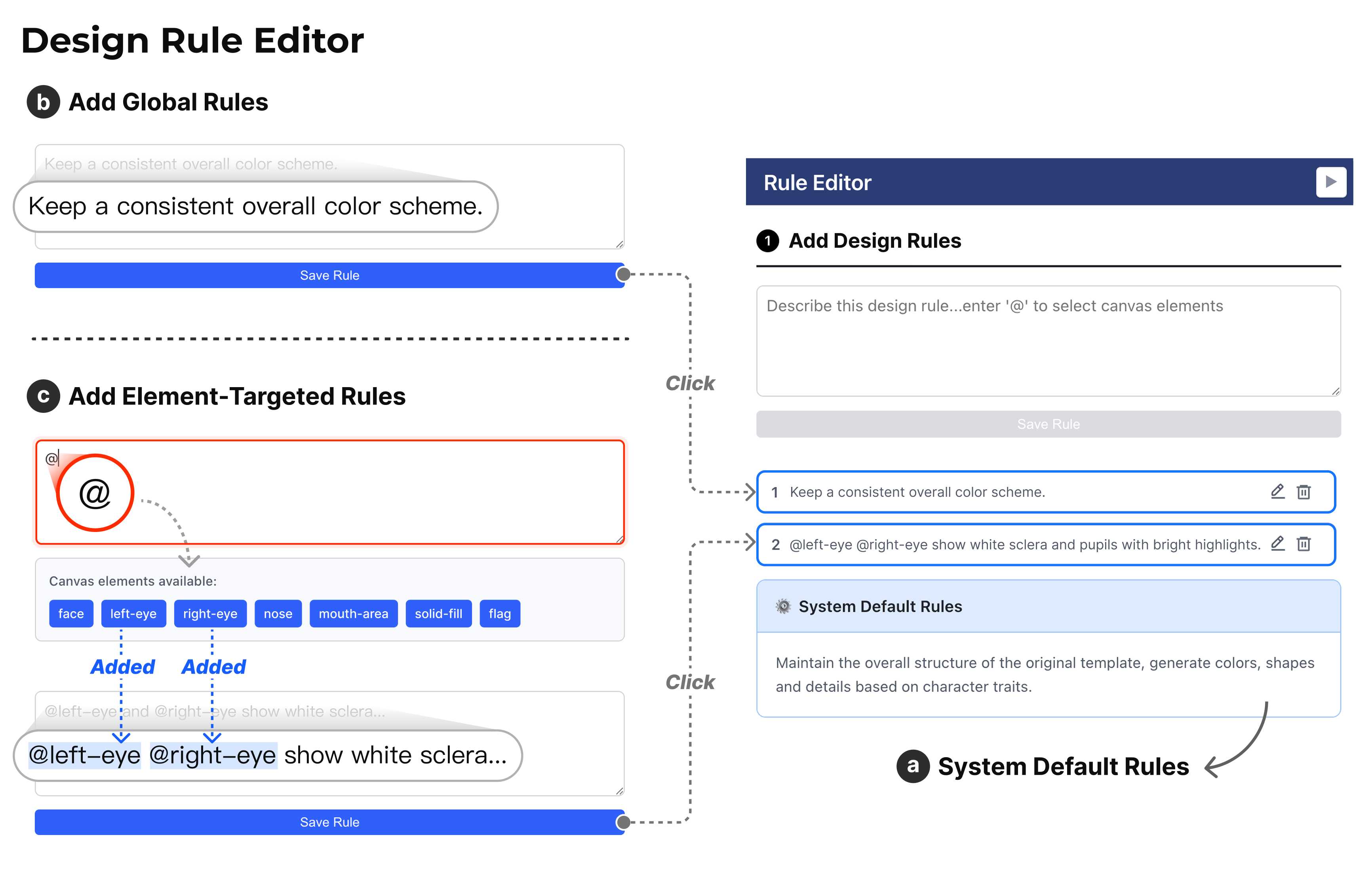}
    \caption{Adding Design Rules: System Default Rules (a) apply when no custom rules are added. Users can override these by adding global rules (b) or element-targeted rules (c) by typing ``@'' to specify tags.}
    \Description{Screenshot of the design rule editor showing default rules, global rule input, and element-targeted rules via @tag.}
    \label{fig:constraint}
\end{figure}

While the Semantic Tag Editor establishes element-level defaults, the Design Rule Editor adds or overrides custom constraints that fill \textbf{[Design Rules]}. If unspecified, system defaults apply (Figure \ref{fig:constraint}-a, \textbf{DC2}). Designers can add rules globally (Figure \ref{fig:constraint}-b) or target specific elements via ``@tag'' (Figure \ref{fig:constraint}-c, \textbf{DC3}). Supported constraints include \textit{style} (color, texture, animation), \textit{detail} (shape, relative scale, distinctive details), and \textit{associated features} (character traits, emotion categories).

\subsubsection{\textbf{Context Mapping Rule Editor (Figure~\ref{fig:overview}-D2).}} 

The Context Mapping Rule Editor defines \textbf{[Context Rules]} by mapping contextual inputs to expression updates. Designers may use simple input forms---such as character descriptions in the personalized face generation phase---or define richer contextual conditions via Define Context (Figure \ref{fig:test}-a, \textbf{DC3}). Designers can then specify contextual factors by selecting from presets (Figure \ref{fig:test}-b) or by creating new ones via Add Context (Figure \ref{fig:test}-c), and compose context-character/context--expression mapping rules (Figure \ref{fig:test}-d) that explicitly link contextual factors to generative behaviors.

\subsubsection{\textbf{Runtime Expression Simulator (Figure~\ref{fig:overview}-E)}} 
\begin{figure*}
    \centering
    \includegraphics[width=\linewidth]{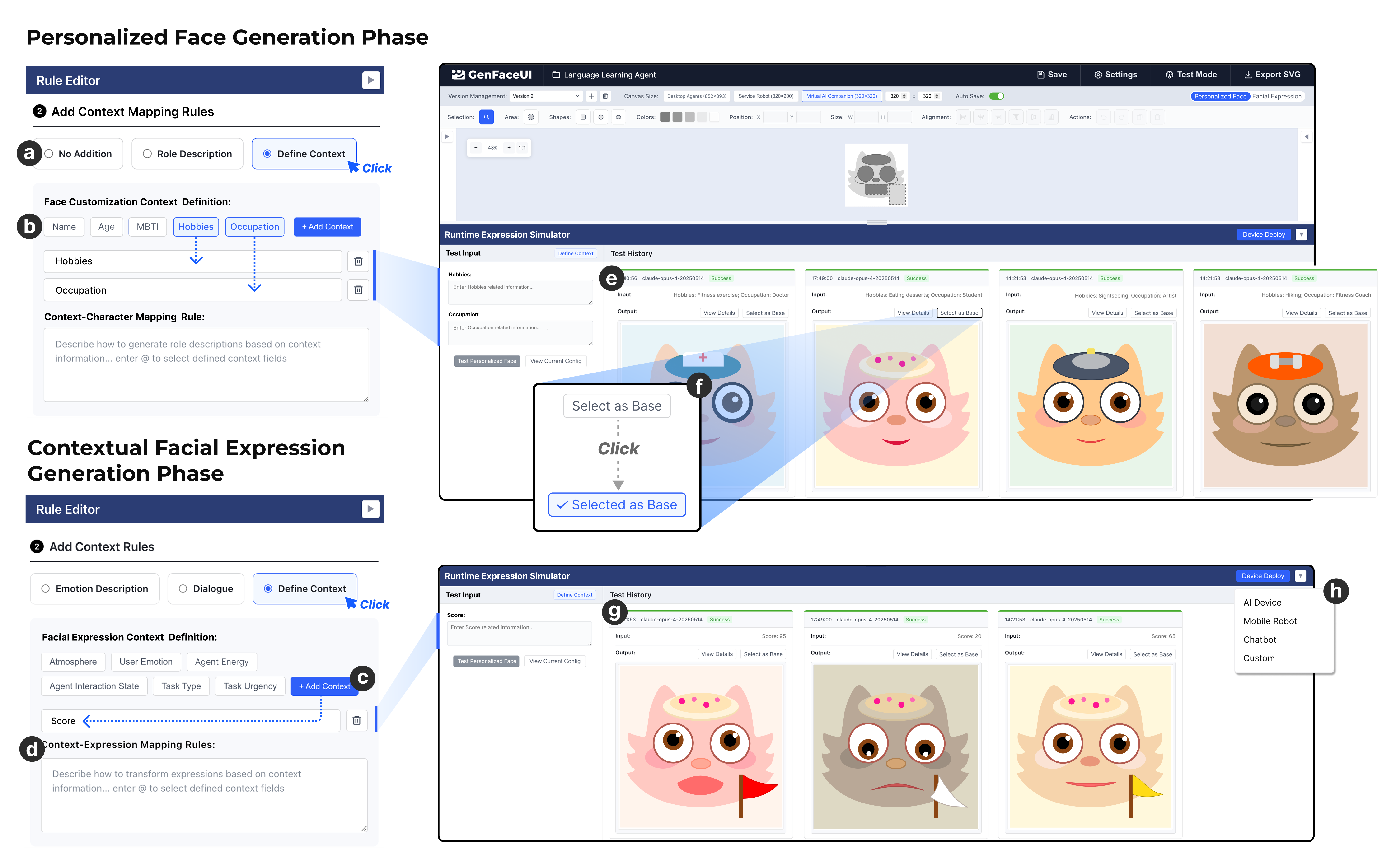}
    \caption{Context Mapping Rule Editor and Runtime Expression Simulator: In the personalized face generation phase, designers define contextual factors and compose mapping rules (a--d). The Runtime Expression Simulator allows designers to input contexts, inspect generated results, and select a base expression as the foundation for contextual variations (e--g). Finally, designs can be deployed to devices (h) for validation.}
    \Description{Composite figure: left panels show defining context factors and mapping rules; right panels show the runtime simulator where inputs are entered and generated facial expressions are previewed and selected.}
    \label{fig:test}
\end{figure*}

The Runtime Expression Simulator in test mode enables designers to verify whether the meta-design templates and defined rules ensure consistency and controllability in generative outcomes (\textbf{DC4}). Designers simulate inputs (Figure~\ref{fig:overview}-E1) for the contextual factors defined in the Context Mapping Rule Editor and review the generated results (Figure~\ref{fig:overview}-E2). In the personalized face generation phase, designers click \texttt{Select as Base} to set a satisfactory outcome (Figure \ref{fig:test}-f). This base then serves as a consistent foundation in the contextual facial expression generation phase. Once results in the Runtime Expression Simulator are broadly satisfactory, designers can click the \texttt{Device Deploy} button (Figure~\ref{fig:overview}-E3, Figure \ref{fig:test}-h) to deploy the workflow to physical terminals for validation in realistic scenarios.

\subsection{Example Walkthrough}
To illustrate the features and workflow of \textsc{GenFaceUI}, this walkthrough shows how designers design generative facial expression for a language-learning agent. We use ``quoted text'' for typed inputs and \texttt{teletype} for UI buttons. Figures~\ref{fig:overview}--\ref{fig:test} visualize the process, and all outputs are generated by \textsc{GenFaceUI}.

\textbf{\textit{Step 1. Sketch the Face Template on the Canvas.}}
In the \textit{personalized face generation} phase, the designer composes facial elements on the \textbf{Face Template Canvas}: two circles for eyes and one ellipse for the nose with \texttt{Shape Drawing}, an area for the mouth and an area for the background fill with \texttt{Area Drawing}, plus an SVG contour imported from Figma\footnote{\url{https://www.figma.com/}} for a fox-like face. To support personalization and context, the designer adds an area beside the face for a learning-related flag and an ellipse for a role-related hat.

\textbf{\textit{Step 2. Add Semantic Tags.}}
The designer tags each element in the \textbf{Semantic Tag Editor}. For example, the designer selects the left eye, clicks \texttt{Organ}, and assigns \texttt{left-eye}; the system shows the auto-added pre-defined rule. The designer repeats for \texttt{right-eye}, \texttt{nose}, and \texttt{mouth} (\texttt{Organ}) and for \texttt{hat} and \texttt{solid-fill} (\texttt{Decoration}). For the flag area, the designer chooses \texttt{Custom} under \texttt{Context} and enters ``flag''.

\textbf{\textit{Step 3. Develop Personalized Face Rules.}}
The designer uses the \textbf{Design Rule Editor} to define constraints for personalized face generation. Example rules: (1) ``@flag renders a waving flag with a brown pole and flag surface; show only when the face changes''; (2) ``@left-eye and @right-eye show white sclera and pupils with bright highlights''; (3) ``Keep a consistent overall color scheme.'' The designer clicks \texttt{Save Rule}. Then, in the \textbf{Context Mapping Rule Editor}, the designer clicks \texttt{Define Context}, selects \texttt{Hobbies} and \texttt{Occupation}, and adds a mapping: ``Face/background fill varies with personality (e.g., brighter for extroverts); the hat's style/color varies with hobbies.''

The designer switches to test mode and iterates until satisfied (Figure~\ref{fig:test}-e). The designer clicks \texttt{Select as Base} to fix the result (e.g., ``Hobbies: eating desserts'', ``Occupation: student'') as the foundation for the next capability.

\textbf{\textit{Step 4. Develop Contextual Facial Expression Rules.}}
The designer switches to the \textit{contextual facial expression generation} phase and adds a custom context factor ``Score'' to represent exam performance. Example mappings: (1) ``Score controls flag state: 80--100 bright red/strong wave; 60--79 steady yellow; 1--59 drooping white''; (2) ``Mouth, pupils, and background fill vary with score.'' The designer tests different scores and inspects generated contextual facial expressions (Figure~\ref{fig:test}-g).

\textbf{\textit{Step 5. Deployment and Validation.}}
The designer exports rules and templates, clicks \texttt{Device Deploy}, and integrates them into the language-learning workflow. The designer then simulates scenarios to observe both personalized face and contextual facial expressions, refining rules iteratively.

\section{Designer Study}
To investigate designers' experiences, perceptions, and expectations regarding the GPFEI framework and meta-design practices, we conducted a qualitative study. \textsc{GenFaceUI} was used as an exploratory prototype to elicit designers' practices and reflections. Based on these goals, we formulated the following research questions:

\begin{itemize}
  \item \textbf{RQ1}: How do designers perceive \textsc{GenFaceUI} in terms of \textbf{consistency, controllability, design agency, and creative support}?
  \item \textbf{RQ2}: What are designers' attitudes toward and expectations of the \textbf{generative personalized facial expression interface} framework?
  \item \textbf{RQ3}: How do designers understand the role of \textbf{meta-design in shaping generative interfaces}?
\end{itemize}

\subsection{Participants}

We recruited 12 designers ($N=12$; $M_{\text{age}}=25.3$ years, $SD=3.5$) through social networks and snowball referrals. Participants first completed a brief screening questionnaire collecting demographic information and prior design experience, which we used to ensure that all participants had relevant design backgrounds. None of the participants had prior exposure to our system.

Participants comprised 8 female and 4 male designers. Six participants had $\geq$3 years of work experience. All participants were interaction designers, with backgrounds spanning product design, HMI, graphic/visual design, AI product design, and human--AI interaction research. All reported using LLM or AI-generated content (AIGC) tools in their work; three (P1, P7, P12) also had experience with vibe coding\footnote{\url{https://x.com/karpathy/status/1886192184808149383}}
.

Before the study, all participants were informed about the study procedures and data collection practices and provided consent in accordance with institutional research ethics and data protection guidelines. Table~\ref{tab:participants-chi} summarizes demographics and prior experience.

\begin{table*}[t]
\footnotesize
\centering
\caption{Participant demographics and AI tool usage (N=12). Data collected via a recruitment questionnaire.}
\label{tab:participants-chi}
\renewcommand{\arraystretch}{1.3}
\setlength{\tabcolsep}{4pt}
\begin{tabular*}{\textwidth}{@{\extracolsep{\fill}} l
  r
  c
  p{0.09\textwidth}
  p{0.37\textwidth}
  p{0.17\textwidth}
  p{0.13\textwidth}
  @{}}
\toprule
\textbf{ID} & {\textbf{Age}} & \textbf{Gender} & \textbf{Work exp.} & \textbf{Design Field and Related Experience} & \textbf{AI tools used.} & \textbf{AI tool use freq.} \\
\midrule
P1  & 23 & F & 3--5 years & Interaction Designer; Human--AI Interaction research & LLM, AIGC, Vibe Coding & Frequent \\
P2  & 21 & F & 1--3 years & Interaction Designer; LLM-based product design & LLM, AIGC & Almost daily \\
P3  & 23 & F & 1--3 years & Interaction Designer; Graphic Design & LLM, AIGC & Almost daily \\
P4  & 24 & M & 3--5 years & Interaction/Product Designer; Human--AI Interaction research & LLM, AIGC & Almost daily \\
P5  & 30 & F & 3--5 years & Interaction Designer; AI Product Designer; Graphic/Illustration & LLM, AIGC & Almost daily \\
P6  & 24 & F & 0--1 years & Interaction Designer; LLM-based product design & LLM, AIGC & Frequent \\
P7  & 28 & F & 5 years & Interaction Designer; AIGC; AI Product Developer & LLM, AIGC, Vibe Coding & Frequent \\
P8  & 23 & F & 0--1 years & Product Design & LLM, AIGC & Almost daily \\
P9  & 21 & M & 1--3 years & AI Product Designer; LLM-based product design & LLM, AIGC & Frequent \\
P10 & 27 & M & 1--3 years & HMI Designer & LLM, AIGC & Almost daily \\
P11 & 30 & F & 5 years & Interaction Designer; AIGC Artist & LLM, AIGC & Almost daily \\
P12 & 30 & M & 5 years & Interaction Designer; AI Product Developer & LLM, AIGC, Vibe Coding & Almost daily \\
\bottomrule
\end{tabular*}
\end{table*}

\subsection{Set up}

Study sessions were conducted in person in a study room. \textsc{GenFaceUI} ran on a desktop computer; participants were provided pen and paper for sketching and reference materials (key concept summaries, task descriptions, exemplar facial expression images). A video camera recorded on-screen and physical interactions, and system audio was captured throughout each session. Figure~\ref{fig:setup}.a summarizes the setup and target deployments.

\subsection{Tasks and Deployment Targets}
We set three facial expression interface design tasks and, for each, provided deployment to a working application so that designers could evaluate behavior in situ (Figure~\ref{fig:setup}-a).

\begin{enumerate}
\item \textbf{Desktop Conversational Agent (Simplest).} Designers created a basic face for a chatbot. The design was deployed to a smartphone placed horizontally on a stand to simulate a desktop robot. After pressing \texttt{Deploy to Device} in \textsc{GenFaceUI}, 
the phone rendered the design and supported live ASR/TTS so that expressions updated with dialogue. This task familiarized participants with the end-to-end workflow.
\item \textbf{Mobile Service Robot.} Designers implemented character customization for a facial expression interface while following core UI principles of aesthetics, consistency, and usability. A \textit{Temi} mobile robot\footnote{\url{https://www.robotemi.com/}} displayed a \textit{Customize Character} control; entering a character description regenerated the on-screen face. This task highlighted the generative nature of the interface and corresponding design strategies.
\item \textbf{Virtual AI Companion (Most Complex).} Designers specified a coherent companion identity from profile input and authored context--expression mapping rules so that expressions adapted to interaction functions. Outcomes were evaluated in a prebuilt app demo with a registration flow (profile input and character generation) and mocked API integration, enabling end-to-end testing. In addition, we provided predefined context data schemas for both face generation and facial expression variation that designers could select to map context information to role descriptions and expression requirements. Example fields included user-profile attributes (e.g., MBTI, age, personality, hobbies) and system context (e.g., environment data, user state, interaction signals), with support for simulated testing.
\end{enumerate}

All tasks were framed as concrete design briefs, with outcomes validated on their target deployments.

\subsection{Procedure}
\begin{figure*}
  \centering
  \includegraphics[width=\linewidth]{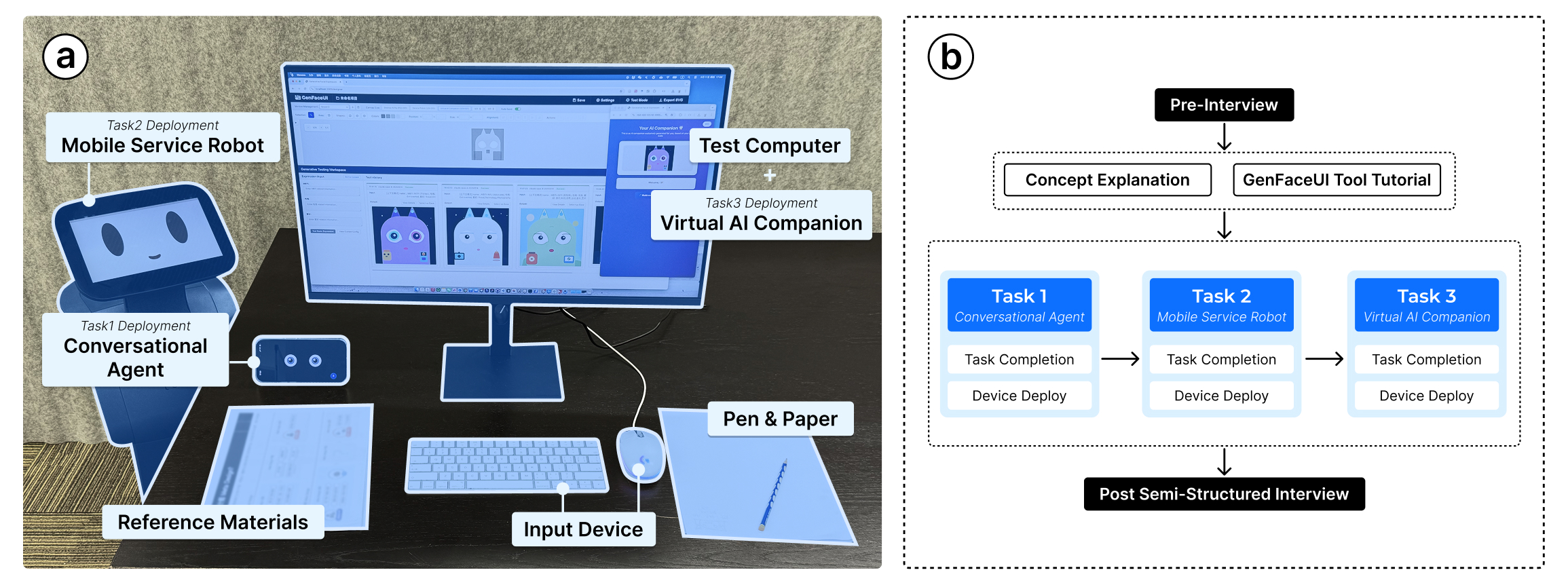}
  \caption{Study setup and target deployments for the three tasks.}
  \Description{The figure shows three deployment scenarios corresponding to the experimental tasks. On the left is a small yellow desktop device representing the AI toy. In the middle is a service robot with a display showing cartoon-style eyes and glasses. On the right is a green cartoon-like character face with expressive eyes and a decorated hat, representing the virtual AI companion.}
  \label{fig:setup}
\end{figure*}

Each session followed a consistent flow. Participants first completed a brief pre-brief and interview covering study motivation, GPFEI concepts, and background information. A tutorial then introduced \textsc{GenFaceUI} through a walkthrough and short free exploration. Participants subsequently worked on the three tasks (Figure~\ref{fig:setup}-b), thinking aloud and requesting assistance when needed. After each task they participated in a brief post-task interview to capture immediate impressions. Finally, participants completed a short wrap-up semi-structured interview. Each session lasted approximately 120 minutes in total. A task was considered complete once the deployed behaviour met the requirements, and participants could refine their designs within the allotted time.

\subsection{Data Collection and Analysis}

We collected three sources of qualitative evidence: system logs, interviews, and think-aloud reflections during the tasks.
\textit{System logs} captured interaction and testing activities, while \textit{interviews} (pre/post) elicited perceptions, attitudes, and practices. \textit{Think-aloud} reflections were noted to capture immediate reasoning.

Our analysis used inductive thematic analysis to address the research questions introduced at the beginning of this section. Two researchers independently coded transcripts and notes, then consolidated codes and resolved disagreements through discussion. Initial open codes were iteratively clustered into higher-level categories, which were refined through constant comparison across participants and tasks. These categories informed the final themes reported in the Results section.
System logs served as contextual evidence linking observed interaction patterns with reported experiences, supporting interpretation of how designers perceived \textsc{GenFaceUI}, the GPFEI framework, and meta-design, and also enabled quantitative summaries of task engagement (e.g., counts of elements created, rules authored, and test generations) reported in the Overall Results.

\section{Findings}

\begin{figure*}
    \centering
    \includegraphics[width=0.9\linewidth]{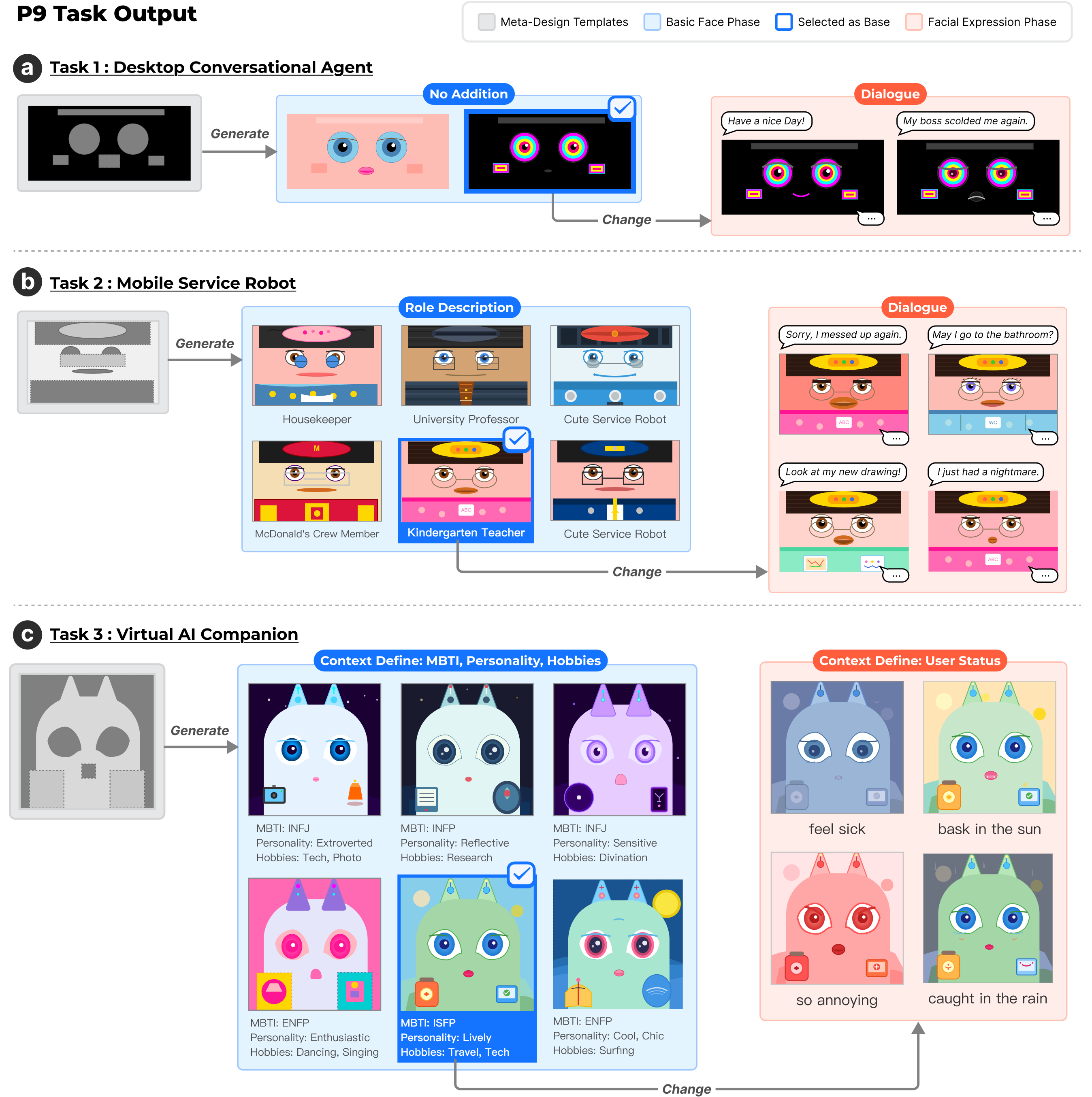}
    \caption{Tested generated results from P9 across three tasks.}
    \Description{Three panels showing Participant P9's generated outputs for Task 1, Task 2, and Task 3, illustrating base faces and contextual facial expressions.}
    \label{fig:P9}
\end{figure*}

\subsection{Overall Results}
Participants' testing behaviors in \textsc{GenFaceUI} revealed active engagement across all tasks. In total, they generated 268 test face, including 202 face generations and 66 facial expression generations. They created 236 interface elements and authored 123 rules overall. 
Engagement patterns were consistent across tasks: in Task1, participants produced on average 5.9 elements and 2.4 rules per participant; in Task2, averages increased to 7.3 elements and 3.8 rules; and in Task~3, they produced 6.4 elements and 4.1 rules.  
Participants also defined 38 custom semantic tags in total, averaging 3.5 per participant. Regarding element usage, all participants who created elements included eye components; ``emoji-area'' was the most frequently used contextual element (9 instances), while ``hat-area'' was the most common decorative element (7 instances). For a representative illustration across the three tasks, see Figure~\ref{fig:P9} (Participant P9); additional examples appear in Appendix~\ref{sec:examples-appendix}.

\subsection{RQ1: How do designers perceive \textsc{GenFaceUI} in terms of consistency, controllability, design agency, and creative support?}

\subsubsection{\textbf{Consistency: stable style with distinctive characters.}}
Participants overwhelmingly agreed that \textsc{GenFaceUI} maintained a consistent visual language across characters while still preserving character-specific distinctions (11/12). This balance allowed designers to express individuality within a cohesive framework, enabling them to work flexibly under a shared standard. As P8 explained, \textit{``The styles of different characters are consistent, yet you can still see their differences.''} Similarly, P7 observed that \textit{``AI-generated results are stable in both style and in how the rules are applied.''} P10 even likened the set to a POP MART series, \textit{``consistent in style yet distinct in features.''}

\textbf{\textit{Divergent views on alignment with design intent.}} While participants widely recognized this visual coherence, views diverged on whether the generated results aligned with their personal design intent. On the one hand, some participants (5/12) felt the outputs matched their expectations, which increased their confidence in the tool. As P2 noted \textit{``The overall look is close to what I had in mind''}. On the other hand, others (3/12) emphasized the unpredictability of outcomes and the lack of stable expectations. P9 commented, \textit{``Many times I don't know what will come next''} . P8 described the experience as \textit{``like opening a blind box---my mood goes on a bit of a roller coaster.''}

\textbf{\textit{Limitations in accuracy and expressiveness.}} The current outcomes show certain discrepancies with designers' intended expectations (7/12). Several participants noted limitations in \textbf{accuracy and controllability}. P10 (a HMI designer) stated, \textit{``The results deviate from my intent, and the details are too simple.''} Similarly, P6 pointed out that the system could not interpret domain-specific terminology such as ``dopamine'', which limited their ability to convey ideas precisely. Beyond accuracy, participants also highlighted issues of \textbf{visual expressiveness}, emphasizing the need for finer-grained controls and richer details. P5 mentioned subtler adjustments to differentiate characters, and P9 asked for \textit{``some variation in gradients''}. Finally, regarding \textbf{dynamic presentation}, four participants wished to see how expressions and actions would be animated, as P9 suggested, \textit{``It would be best to see how it moves.''} Yet others felt the current animations were unnatural and \textit{``a bit uncanny''} (P6).

\subsubsection{\textbf{Controllability: Need for finer-grained control.}} We found that the participating designers had high expectations for the tool's controllability. From a functional perspective, they especially appreciated that the system offered \textbf{area tools} for targeting specific facial regions and \textbf{semantic tags} for quickly distinguishing element categories. The default addition of semantic rules further reduced unpredictability and helped steer the generated results. As P7 noted, \textit{``semantic tags saved considerable time in defining rules''}.

\textbf{\textit{Challenges in system usability.}} Many described the process as more akin to playful art creation than disciplined design, noting that the interface relies on free-form canvas and text input rather than the structured operations typical of professional tools. A key issue was the \textbf{lack of structured interaction}, since most rules were added through text, making the workflow feel like \textit{``writing prompts'' rather than ``doing design''} (P7), with calls for more direct, design-oriented operations. Participants further criticized \textbf{inefficient operation and workflow support}: P5 noted the need for repeated experimentation to craft effective prompts, and P10 suggested a node-based procedural approach to improve traceability. Finally, participants highlighted \textbf{difficulty with complex tasks}, observing that the system struggles to support highly structured or sophisticated UI projects; as P2 put it, the current prompt model and hierarchy are \textit{``a bit too simple,''} leading to misunderstandings and extra revisions.

\subsubsection{\textbf{Design Agency: Expanding design space.}} All participants agreed that the tool genuinely broadened the possibilities for facial expression interfaces, moving beyond traditional solutions. They noted that it allowed them to \textit{``constrain the expressive scope of generated faces''} (P9) and that \textit{``customizing different characters was very efficient with it''} (P3).

\textbf{\textit{Limited by missing professional authoring features.}} However, our framework still has the following limitations in supporting designers' agency. As P7 noted, the platform relies mainly on a drawing canvas and text input, without detailed editing or layer management,  which making it \textit{``lacks control over fine details or overall comprehensiveness''}(P9). 
Current AI models struggle to understand specialized design terms. P5 said \textit{``its design hierarchy is somewhat simplistic,''} and P6 added that \textit{``its understanding of my prompt is rather simplistic.''} Compared to the powerful capabilities of professional tools, many respondents characterized this generative system as more \textit{``interesting''} or \textit{``novel''} rather than truly useful for detailed design. P8, for example, praised it as \textit{``very interesting... the interaction pages are quite complete,''} but did not believe it could handle meticulous professional design work.

\textbf{\textit{From serendipity to loss of control.}} The initial ``surprise'' from serendipitous AI outputs diminished, instead amplifying designers' sense of lost control. Most participants (7/12) lacked a clear understanding or foresight of the AI's capability boundaries. P3 said, \textit{``It matters whether the result matches my expectation''}, and P5 added, \textit{``If the results keep failing to match, it just feels like arbitrary generation''}. P8 emphasized that the absence of immediate feedback and range previews made control feel like \textit{``pulling a random gacha card''}, and P10 noted that such uncertainty seriously undermined willingness to use the tool. Compared with serendipity, participants prioritized controllable randomness. Several participants (5/12) suggested the system should provide more explanations of how results are produced to better support understanding and customization of the generation mechanism.

\subsubsection{\textbf{Creative Support: Lowers the barrier to creation.}}  
Participants felt that the system significantly lowered the entry barrier for design, opening opportunities even for non-professionals (P2, P8). Beyond accessibility, most also believed it expanded their creative capacity, helping them express and develop ideas in the early, exploratory stages (8/12). As P1 noted, \textit{``It expands your capabilities,''} and P2 added that when \textit{``drawing broadly on inspiration, it can serve to inspire.''} However, some participants expressed uncertainty about whether such inspiration could be sustained throughout the design process, suggesting that the tool's primary value lies in rapid drafting and idea exploration rather than in supporting deeper creativity.

\textbf{\textit{Structured controls to better support creativity.}}  
Participants emphasized the need to move from prompt-writing toward more design-oriented interactions. As P6 put it, \textit{``It should be about designing, not writing prompts,''} and P7 argued for inputs that are \textit{``structured rather than textual descriptions,''} with features such as structured tags, parameter visualization, and key-parameter locking (P5). Others highlighted the importance of transparent parameter feedback to support controllable adjustments across different tasks (P1).

 \subsection{RQ2: What are designers' attitudes toward and expectations of the generative personalized facial expression interface framework?}

Designers regarded the framework with both enthusiasm and caution. While they acknowledged its promise for personalization and contextual expression (12/12), they also emphasized the risks of output control and the need to ground evaluation in real-world contexts.

\textbf{\textit{Optimistic about future applications.}} All participants acknowledged the framework's value for personalization and contextual expression. For personalization, they emphasized both brand-oriented and end-user customization. For instance, P5, leveraging prior graphic design experience, added a bright star in the top-left corner to parody a corporate logo and reinforce brand identity in Task 2. Participants also noted support for end-user customization, such as easing \textit{``content customization and extending to slide design''} (P4), which allowed exploration of styles and editability. Regarding contextual expression, participants highlighted its role in affective interaction, describing it as fostering \textit{``more human-like and scenarized presence''} (P7) and offering \textit{``better emotional feedback''} (P3).

\textbf{\textit{Risks in generative output control.}}  
Participants highlighted the difficulty in achieving reliable control, noting that high-quality results \textit{``still rely on professional oversight''} (P8). Another risk involved the uncontrollability of the generative space, where \textit{``poor results could negatively affect user experience and brand communication''} (P6). Participants also cautioned that granting users unrestricted control introduces unpredictability, underscoring the need for safeguards to prevent unintended results. As P7 explained, \textit{``giving users unrestricted control introduces unpredictability,''} and P10 similarly stressed that \textit{``complete content control is necessary to avoid unintended results.''}

\textbf{\textit{Heavy reliance on real-world context.}} Meta-design results are assessed through their actual use on end-user devices. When engaging in Meta-design, designers should \textit{``develop a deep understanding of the design requirements and challenges of the intended context''}, so they can pre-assess the likely generation outcomes. Some respondents even noted that without sufficient task descriptions or scenario information, they \textit{``have no way to find a direction to add details''} (P9), or that when \textit{``requirements and goals are unclear''}, it is difficult to \textit{``set boundaries''}(P10). Consequently, designers'core competency shifts from emphasizing the delivery of specific design outputs to cultivating a deep understanding of the design challenge.

\subsection{RQ3: How do designers understand the role of meta-design in shaping generative interfaces?}
The study shows that designers understand meta-design not as a fixed concept, but as a practice negotiated through authority, interpretation, and shifting responsibilities in shaping generative interfaces.

\textbf{\textit{Maintaining authority in generation quality.}}  
All participants believed that designers must be involved and be able to control the design quality of generated content (12/12). In Meta-design, designers must still be responsible for aesthetic benchmarks, requirements decomposition, and the final presentation. P3 noted that even with AI assistance, the final visual outcome and design structure still required professional judgment, and P9 argued that in setting aesthetic standards and formulating prompts, \textit{``designers remain the authority''}, and one cannot rely entirely on the generative system. Likewise, P8 emphasized the importance of understanding and unpacking requirements, \textit{``thinking designers need to strengthen their ability to deconstruct a character/image''}.P10 maintained that \textit{``conveying emotional expression through visual means''} remains the designer's responsibility.

\textbf{\textit{Gradual understanding of meta-design.}} During testing, we noticed that participants showed differing interpretations of the design tasks. Some initially treated it as optimizing or well-designing a single character or element. For example, P10 began by enforcing the mouth element not change with facial variation, while P6 specified exact RGB values for the eye element. As for the reason, P9 noted that \textit{``without a concrete scenario, designers must rely on imagination and adapt to contextual variations.''} Others more quickly recognized that the task was not merely to \textit{``build an interface''} (P2, P8, P12), but to construct reusable rules. P3 said she came to \textit{``establish the constraints of a design specification''} after completing the Task 2.

\textbf{\textit{From artifact creators to rule-makers.}}  
Designers' responsibilities are shifting from crafting specific design artifacts to constructing processes and rules that govern generation outcomes. P7 stressed the need to lower the barrier for end-user participation by offering customizable parameters, noting that \textit{``it's no longer just a fixed, manufacturer-set design; instead, everyone can join this loop... to define the final outcome,''} highlighting the importance of accessible interfaces for participation. Beyond enabling participation, designers also take on new roles that emphasize control and creativity: as \textit{``guardians''} (P5), they keep styles consistent and avoid improper results; as \textit{``translators''} (P1, P3), they turn goals and needs into clear rules and exercise judgment to improve outcomes; and as \textit{``explorers''} (P8), they manage the surprises of generated results while \textit{``keeping space for creativity beyond fixed solutions''} (P4). This illustrates a broader shift from designers as artifact creators to system designers and rule-makers.

\section{Discussion}

In this work, we examined generative facial expression interfaces through a meta-design lens, integrating a conceptual framework, a proof-of-concept tool, and a qualitative study. Designers evaluated the approach positively, noting clearer controllability, stronger stylistic coherence, and greater creative support, while also highlighting challenges in expressiveness, predictability, and interpretability. Taken together, these findings consolidate this work's contributions at three levels: system-level insights into how \textsc{GenFaceUI} materializes meta-design in practice, framework-level implications for the structure and utility of GPFEI, and broader conceptual advances in understanding meta-design as a foundation for designer--AI collaboration in generative interfaces. The discussion builds on these conclusions by first distilling lessons from the tool, then articulating implications for the framework, and finally situating meta-design as an emerging perspective for future generative interface practice.

\subsection{Design Lessons from GenFaceUI}

\textbf{Meta-design tools for generative facial expression interfaces should establish a consistent human--AI collaboration model across design time and run time.} Designers in our study expected co-creation within the tool, not merely a division where designers specify at design time and the AI generates at run time. Participants often described difficulty anticipating how their rules would manifest in generated expressions, which reflects the uncertainty and trial-and-error repeatedly reported in our results and underscores the importance of \textbf{making AI behavior observable at design time} to support reliable mental models and clarify responsibility (Lesson 1). Making the AI a visible ``design actor'' at design time---providing immediate feedback on rule effects, brief explanations for generative behavior and failure cases, and constraint-aware suggestions---clarifies responsibilities and supports accurate mental models for deployment, aligning with views of generative interfaces as artifacts ``created, refined, or evolved'' through computational processes \cite{lee_towards_2025}. Such expectations resonate with established perspectives on mixed-initiative interaction \cite{horvitz_principles_1999} and the long-standing view of computers as partners in the creative process  \cite{lubart_how_2005}, while extending meta-design principles of distributing agency across humans and AI \cite{barricelli_advancing_2024}. Our findings further indicate that facial expression interfaces, which require subtle semantic and stylistic coherence, amplify designers' need to observe the AI's behavior during rule authoring in order to calibrate trust and preview expressive variation.

\textbf{Maintaining this collaboration requires tools that help designers structure intent predictably, interpret system behavior transparently, and work with richer forms of specification.} 
Our findings show that designers often struggled to anticipate how rules would interact or override one another, and they expressed a need for more predictable ways to shape generative outcomes. This indicates the importance of \textbf{providing structured control} through explicit parameters and visible constraint relationships (Lesson 2). Designers also frequently reported uncertainty about why unexpected expressions were produced and requested clearer indications of how the system interpreted their rules. This highlights the value of \textbf{offering transparent feedback} such as instant previews and brief explanations of rule activation (Lesson 3). Finally, many designers found text-only rules limiting. They wanted ways to express stylistic intent, visual reference, or semantic nuance without relying solely on prompt-like formulations, motivating support for \textbf{supporting richer specification} through exemplar-based inputs or guided rule construction (Lesson 4).
These mechanisms echo broader guidelines for human--AI interaction that highlight feedback, intelligibility, and error handling as central to building usable and trustworthy co-creative systems \cite{amershi_guidelines_2019}, and help ensure alignment between design-time intent and run-time behavior.

\subsection{Generative Facial Expression Interfaces in Intelligent Agents}

The GPFEI framework repositions facial expression interfaces as functional communicative infrastructure for a wide range of intelligent agents across virtual agents, avatars, game NPCs, and other virtual characters, as well as embodied robots. Designers in our study described generative expressions as supporting the communication of agent identity, internal state, intent, and social cues across contexts, indicating that GPFEI provide a unified expressive mechanism that can operate across diverse embodiments and system architectures. This expands the role of facial expressions beyond reactive affect displays, helping agents mitigate the language-dominant nature of LLM-driven interaction and enabling more coherent multi-modal personas \cite{ben_youssef_towards_2015,gao_learning_2022}. Under this view, GPFEI do not merely enlarge a visual design space; they suggest a pathway for how expressive behavior might serve as an interaction layer across virtual and embodied agents, potentially contributing to richer and more balanced human--agent interaction.

At the same time, the limitations of \textsc{GenFaceUI} indicate that the current instantiation of GPFEI still operates primarily as a generative visual interface rather than a fully integrated communicative layer within agent systems. Participants consistently noted missing temporal continuity (e.g., smooth transitions, interpolation) and incomplete multimodal coordination with voice, gesture, or contextual cues. These constraints reduce the functional role of expressions in situated interaction and risk confining generative faces to visual mockups \cite{wang_design_2022,ebisu_see_2025}. Importantly, these limitations are not only technical but conceptual: realizing GPFEI's full potential requires integrating expression synthesis with agent timing dynamics, task context, and behavioral coordination mechanisms. Addressing continuity and multimodality is therefore critical if GPFEI are to evolve into reliable, situated communicative channels capable of supporting real-time interaction across both virtual and physically embodied intelligent agents.

\subsection{Meta Design: Designer Agency, Co-Agency with AI, and Role Allocation}

Designing generative interfaces requires a clear division of labor in meta-design. As our study of generative expressive interfaces illustrates, the central question is not only who performs which tasks, but how the underlying conditions for expression are defined. Because expressions are produced dynamically rather than predetermined, designers exercise agency by defining the rules, constraints, and semantic structures that guide what the system can generate. This perspective aligns with prior work describing generative interfaces as distributed across humans and AI systems \cite{lee_towards_2025}, and it clarifies why a meta-design lens was essential in our study. It helped us see why designers in our findings focused less on specifying fixed outputs and more on shaping the conditions under which expressive outcomes could emerge.

Generativity also introduces a practical form of co-agency between designers and the AI. Designers articulate intent through templates, tags, and rules, while the AI interprets these inputs at run time and adapts them to situational context. Our findings show that expressive outcomes were shaped jointly by designer specifications and the system's run-time interpretations, rather than being determined solely by what designers prescribed. Designers therefore evaluated both the final expression and the reasoning process behind it, indicating that understanding how the AI interprets rules is a core part of their work. This interdependence demonstrates that co-agency is inherent to generative interfaces rather than a secondary layer added after rule authoring.

Because expressive behavior arises from multiple contributors, role allocation becomes an essential part of meta-design. Clarity is needed about who defines expressive intent, who interprets it in context, and who is responsible for evaluating appropriateness. Our findings show that designers were particularly sensitive to these boundaries when system outputs diverged from their expectations or carried social implications. Making these roles explicit supports creative flexibility while ensuring responsible deployment, and it highlights how a meta-design perspective can help structure generative expressive behavior in intelligent agents, including their multimodal interaction capabilities, as well as other emerging forms of generative user interfaces.

\section{Limitations \& Future Work}

\subsection{Limitations}
Building on the insights from our designer study, several limitations remain when considering the broader applicability of GPFEI as an expressive interface for intelligent agents. At the framework level, GPFEI remains an early-stage, design-time--oriented approach that does not yet realize several capabilities expected of mature facial expression interfaces. Although the framework clarifies how designers formulate rules and shape generative spaces, it does not support continuous expressive transitions, finer-grained modulation, or adaptive responses to context. These gaps limit its immediate suitability as a functional expressive layer and highlight the need to extend the framework toward behaviors that unfold over time and coordinate with other modalities.

From the perspective of designer experience, the current system exhibits two distinct limitations. First, it provides only partial support for fine-grained and professional-level control. Designers lacked mechanisms for specifying detailed visual parameters, manipulating complex structures, or achieving precise stylistic adjustments, which limited the expressive fidelity they could realize through rule authoring. Second, the system's reliance on LLM-driven interpretation introduces instability into the generation process. Even when designers articulated clear rules, the model did not consistently apply or interpret those specifications, and similar inputs could yield noticeably different outputs. This unpredictability complicates designer decision-making and also affects the credibility of our study observations, as designer feedback necessarily reflects interaction with a system whose behavior is not fully stable.

Technical limitations further shape the scope of the prototype. The SVG-based pipeline demonstrates the feasibility of generating expressions through rule-driven vector construction, but it is not intended as a long-term technical solution. Its current implementation serves primarily as a research probe, and more capable generative models will be needed for future systems that require higher fidelity, adaptability, or temporal sophistication.

Finally, our empirical findings derive from a designer-focused study rather than evaluations with final users or situated human--agent interaction. The study reveals how designers engage with meta-design processes but does not demonstrate how generated expressions influence usability, trust, engagement, or communicative clarity for end users. The absence of quantitative measures further limits the generalizability of our conclusions. Together, these limitations indicate that while this work establishes a conceptual and methodological foundation for generative expressive interfaces, substantial steps remain before GPFEI can be validated as an expressive infrastructure for real-world intelligent-agent interaction.

\subsection{Future Work}

Building on the GPFEI framework and the insights from our designer study, several directions can advance this work toward more mature expressive interfaces for intelligent agents.

\textbf{1. Advancing expressive capabilities within the GPFEI framework.}
Future work can extend the expressive repertoire of GPFEI to better support the characteristics expected of functional facial interfaces. Key directions include enhancing visual dynamics, enabling smoother transitions across expressions, and improving coordination between facial behavior, linguistic output, and other communicative cues. These improvements are essential for achieving coherent real-time expression and for enabling future in-the-wild evaluation of expressive intelligent agents.

\textbf{2. Strengthening professional design support in \textsc{GenFaceUI}.}
Although \textsc{GenFaceUI} validates the GPFEI framework, its current implementation lacks many professional-level functions required for detailed visual or motion design. Adding more precise control mechanisms, richer editing affordances, and higher-fidelity expression specification tools would allow designers to articulate expressive intent more accurately and work at a level closer to their existing design practices.

\textbf{3. Evolving \textsc{GenFaceUI} into a collaborative meta-design environment.}
Beyond adding features, the tool itself can shift from a static editor to a collaborative space where the AI actively participates throughout the design process. This includes interpreting designer intent, clarifying rule interactions, providing explanations, and helping maintain continuity between design-time rules and run-time behavior. Supporting such co-agency would help designers build clearer mental models of the AI and more effectively guide and constrain how the interface changes during generation.

\subsubsection{\textbf{Broader Research Directions}}
Beyond the immediate development of GPFEI and \textsc{GenFaceUI}, our work points to broader research directions that concern the application of generative expressive interfaces and the further articulation of meta-design within intelligent agent systems.

\textbf{1. Application-oriented research directions.}
GPFEI offers opportunities to design customizable and context-aware expressive agents across diverse domains. Potential applications include:
\begin{itemize}
\item generative characters and non-player agents in games,
\item telepresence and communication avatars used in remote collaboration,
\item task-oriented virtual agents such as virtual clinicians, counselors, or educators, and
\item embodied agents including social robots and humanoid systems.
\end{itemize}

These scenarios require expressive behavior that is multimodally coordinated, aligned with situational norms, and responsive to user needs. A central research challenge lies in supporting personalization and adaptivity while maintaining stability, controllability, and safety. Extending GPFEI into these contexts would not only broaden its applicability but also provide concrete opportunities to evaluate how generative facial expression interfaces, as well as broader multimodal expressive behaviors, perform under real-world constraints, including how users interpret, respond to, and rely on such behavior.

\textbf{2. Theoretical and meta-design research directions.}
Future work may extend and refine meta-design as a foundation for generative user interfaces. One line of inquiry concerns how meta-design practices generalize across different interaction settings and how they can be operationalized and validated in a broader range of generative UI scenarios. This includes examining how design-time intent and constraint-setting translate into run-time behavior across diverse applications.

A second line of inquiry concerns the evolving roles and responsibilities of human designers and AI designers as generative interfaces move into real-world deployment. Future work may investigate how both human and machine contributors participate in shaping generative behavior, how design authority and interpretive responsibility should be allocated between them, and how other stakeholders such as developers or domain experts engage in this process. These questions also point to the need for design systems built for generative agents as primary users of design artifacts rather than for human operators. Advancing this line of research will be essential for integrating meta-design principles into generative user interfaces, enabling greater consistency, controllability, and safety as these systems become more widely deployed.

\section{Conclusion}

This paper examined generative facial expression interfaces for intelligent agents, where run-time synthesis raises practical challenges of control, coherence, and alignment with design intent. Adopting a meta-design lens, we proposed a conceptual framework (GPFEI) of generative facial expression for intelligent agents, implemented it in the \textsc{GenFaceUI} tool, and used a designer study to explore how rule-bounded spaces, character identity, and context--expression mapping shape the design process. The findings indicate gains in controllability and consistency, alongside needs for structured visual mechanisms, lightweight explanations, and clearer role allocation between designers, AI, and other stakeholders. Overall, the work advances understanding of generative facial expression interfaces and outlines directions for connecting design-time specification with situated user experience, contributing to the broader design paradigm of generative interfaces.

\begin{acks}
This work was supported by the Research Start-up Fund (Grant No.\ Y01656116) and the Research Start-up Matching Fund (Grant No.\ Y01656216) from the Southern University of Science and Technology. It was also supported by the Opening Project of the State Key Laboratory of General Artificial Intelligence, BIGAI/Peking University, Beijing, China (Project No.\ SKLAGI2025OP14), the Tongji University National Artificial Intelligence Production-Education Integration Innovation Project, the Fundamental Research Funds for the Central Universities from Tongji University, and the Shanghai Gaofeng Project for University Academic Program Development.
\end{acks}

\bibliographystyle{ACM-Reference-Format}
\bibliography{references}

\appendix

\section{Semantic Rules}
\label{app:semantic}
\textbf{Semantic Rules in Personalized Face Generation Phase}
\begin{lstlisting}[breaklines=true, breakatwhitespace=true]
"face": "Draw the face area based on [template-face] with a suitable skin tone matching the character's traits. Add appropriate facial features such as beard, blush, or dimples. Do not use pure white (#ffffff) or the original template colors. Keep the overall contour and position unchanged, but allow detail elements (e.g., eyelashes, eyebrows) to extend beyond the contour. To better shape the character, you may add accessories such as glasses."
"left-eye": "Draw the left eye based on [template-lefteye], keeping detailed structures such as eye socket, eyeball, pupil, and white highlights. Add character-specific details if appropriate, such as drawing eyebrows that do not obstruct the eye area. Do not draw eyelashes. Keep contour and position unchanged."
"right-eye": "Draw the right eye based on [template-righteye], keeping detailed structures such as eye socket, eyeball, pupil, and white highlights. Add character-specific details if appropriate, such as drawing eyebrows that do not obstruct the eye area. Do not draw eyelashes. Ensure the eye stays within the designated area, with contour and position unchanged."
"nose": "Draw a cartoon nose based on [template-nose], adding detail features according to character traits. Remove the original shape outline. Ensure the nose stays within the designated area, with contour and position unchanged."
"mouth": "Draw a cartoon mouth based on [template-mouth], adding details such as shape, size, color, lipstick, or accessories depending on character traits. Remove the original shape outline. Ensure the mouth stays within the designated area, with contour and position unchanged."
"left-ear": "Draw the left ear based on [template-leftear]. Remove the original shape outline. Ensure the ear stays within the designated area, with contour and position unchanged."
"right-ear": "Draw the right ear based on [template-rightear]. Remove the original shape outline. Ensure the ear stays within the designated area, with contour and position unchanged."

"scene": "Draw a background scene in [template-scene] that matches the character's traits. It should include environmental details, fill the entire area without exceeding the boundary, and preserve the contour. Do not directly use the default template colors."
"scene-item": "Draw an object from the character's scene based on [template-sceneitem]. Keep contour and position unchanged."
"emoji": "Preserve [template-emoji] but render it with no color, keeping it invisible during character generation."
"text": "Preserve [template-text] but render it with no color, keeping it invisible during character generation."
"clothing": "Draw clothing based on [template-clothing], adding detail elements that reflect the character's traits. Remove the original shape outline. Keep contour and position unchanged."
"hat": "Draw a distinctive hat based on [template-hat] that matches the character's traits and style. Add detail features as appropriate. Remove the original shape outline. Keep contour and position unchanged."
"jewellery": "Draw accessories based on [template-jewellery] that suit the character, adding detail features as appropriate. Remove the original shape outline. Keep contour and position unchanged."
"color": "Draw a solid color block based on [template-color]. Do not use the original template color or draw any border. Select colors intelligently according to character traits. Keep contour and position unchanged."
\end{lstlisting}

\textbf{Semantic Rules in Contextual Facial Expression Generation Phase}
\begin{lstlisting}[breaklines=true, breakatwhitespace=true]
"face": "Express emotions by modifying [template-face], ensuring the overall contour and position remain unchanged."
"left-eye": "Express emotions by modifying [template-lefteye], keeping the detailed structures of the eye unchanged and the color consistent."
"right-eye": "Express emotions by modifying [template-righteye], keeping the detailed structures of the eye unchanged and the color consistent."
"nose": "Express emotions by modifying [template-nose], keeping the detailed structures of the nose unchanged."
"mouth": "Express emotions by modifying [template-mouth], keeping the detailed structures of the mouth unchanged.",
"left-ear": "[template-leftear] is the left ear, with the contour and position kept unchanged."
"right-ear": "[template-rightear] is the right ear, with the contour and position kept unchanged."

"scene": "Change the scene drawn in [template-scene] to express emotions or reflect the interactive context. Fill the entire area without exceeding the boundary."
"scene-item": "Change the scene item drawn in [template-sceneitem] to reflect the interactive context."
"emoji": "Add emoji symbols in the [template-emoji] area to express emotions or reflect the context. Ensure the content is within the area and displayed correctly. You will determine the number, size, and layout of emojis based on the overall layout. If the area is nearly square, add only one; if the area is rectangular or irregularly shaped, multiple emojis can be added, and ensure the layout is reasonable and covers the entire area."
"text": "Change the text content in [template-text] to express emotions or reflect the context. Remove the original shape outline and fill color. You will determine the number of words and the size of the text based on the overall layout. Ensure the text is within the area and the position remains unchanged."
"clothing": "[template-clothing] is clothing, with the contour and position kept unchanged."
"hat": "[template-hat] is a hat, with the contour and position kept unchanged."
"jewellery": "[template-jewellery] is accessories, with the contour and position kept unchanged."
"color": "Express emotions or reflect the context by changing the color of [template-color]. Keep the contour and position unchanged."
\end{lstlisting}

\section{Prompt Template}
\label{app:prompt}

\textbf{Prompt for Personalized Face Generation}
\begin{lstlisting}[breaklines=true, breakatwhitespace=true]
[Role]
You are a cartoon character generator. Your task is to generate a personalized character image in SVG code based on the given character description or character design requirements.

[SVG Template]
${designRulesData.template}

[Design Rules]
${designRulesData.rulesDescription}

[Character Description]
You should generate the character description according to the character design rules and the character design context.

##Character Design Rules
${mappingRules}

##Character Design Context
${contextSections}

[Workflow]
1.First analyze the [Character Description] to understand the requirements of character design.
2.Then, based on the [Template] and [Design Specifications], draw the character image.
3.Output SVG.

[Overall Rules]
1.Only output the raw code itself. Do not add any formatting symbols, code block markers, or any other text.
2.The SVG elements in the [SVG Template] constrain the shapes or regions you can draw. You must draw content strictly within the predefined shapes or regions. It is forbidden to add any elements not defined in the [SVG Template].
3.Within the [Design Specifications], designer-defined rules have higher priority than the basic specifications.
4.It is strictly forbidden to use any color attributes (fill, stroke, etc.) from the [SVG Template]. You must reselect all colors according to the character's traits and context. Template colors are only for locating regions, not for final use. In particular, the face element must use an appropriate skin tone and must not remain white.
5.It is strictly forbidden to add any organs or elements not defined in the [SVG Template]. You may only draw within element regions that have existing IDs in the template. Do not add organs such as mouth, ears, or hair if they are not defined in the template.
6.Ensure all element IDs are preserved.
7.If the design specifications require adding textures, you must render them as solid color blocks rather than curve patterns.
\end{lstlisting}

\textbf{Prompt for Contextual Facial Expression Generation}
\begin{lstlisting}[breaklines=true, breakatwhitespace=true]
[Role]
You are a robot dynamic expression generator. Your task is to generate personalized emotional expressions with emotional features in SVG code based on the expression description (including shaping rules and shaping context).

[Basic Character]
${designRulesData.template}

[Design Specifications]
${designRulesData.rulesDescription}

[Expression Description]
You should generate the expression description according to the expression shaping rules and the expression shaping context.

Expression Shaping Rules
${mappingRules}

Expression Shaping Context
${contextSections}

[Workflow]
1.analyze the [Expression Description] to understand the requirements for shaping the character's emotional expression.
2.based on the [Basic Image] and [Design Specifications], draw personalized emotional expressions.
3.Output SVG.

[Overall Rules]
1.Only output the raw code itself. Do not add any formatting symbols, code block markers, or any other text.
2.Ensure that rich and obvious changes in expression are drawn in accordance with the [Basic Image] and [Design Specifications].
3.Ensure that the image is based on the [Basic Image], and maintain the consistency of the image while changing expressions. Strictly follow the [Design Specifications] to draw the changed content.
4.The SVG elements with ids in the [Basic Image] constrain your drawing area. Please draw content only within the areas of the existing SVG elements.
5.It is strictly forbidden to add any organs or elements not defined in the [Basic Image]. Only the content of the element areas with ids already existing in the template can be drawn, and any elements not defined in the template must not be added.
6.It is strictly forbidden to delete any elements already defined in the [Basic Image]. All defined elements need to be generated again.
7.Among the [Design Specifications], the [Designer - defined Design Specifications] have higher priority than the [Basic Design Specifications].
\end{lstlisting}

\section{Examples of Study Participants' Creative Facial Generative Output}
\label{sec:examples-appendix}

Table~\ref{tab:task1-p1} shows P1's output in Task 1, and Table~\ref{tab:task3-p2} shows P2's output in Task 3.

\aptLtoX{
\begin{table*}[t]
  \centering
  \caption{P1 generates characters with defined organs and context-related dynamic emojis in Task 1.}
  \footnotesize

  \setlength{\tabcolsep}{6pt}
  \renewcommand{\arraystretch}{1.12}

  \begin{tabular}[t]{p{0.58\textwidth} p{0.36\textwidth}}
    \toprule
    \textbf{Rules Definition} & \textbf{Output Visualization} \\
    \midrule

    \multicolumn{2}{l}{\textbf{Semantic Tags}} \\
    \leftpad
    1.\;Preset Tags: [face], [left-eye], [right-eye], [mouth-area], [emoji-area] \lsep\newline
    2.\;Custom Tags: [left-eyebrow-area], [right-eyebrow-area], [left-blush], [right-blush]
    &
\includegraphics[height=\imgH,keepaspectratio]{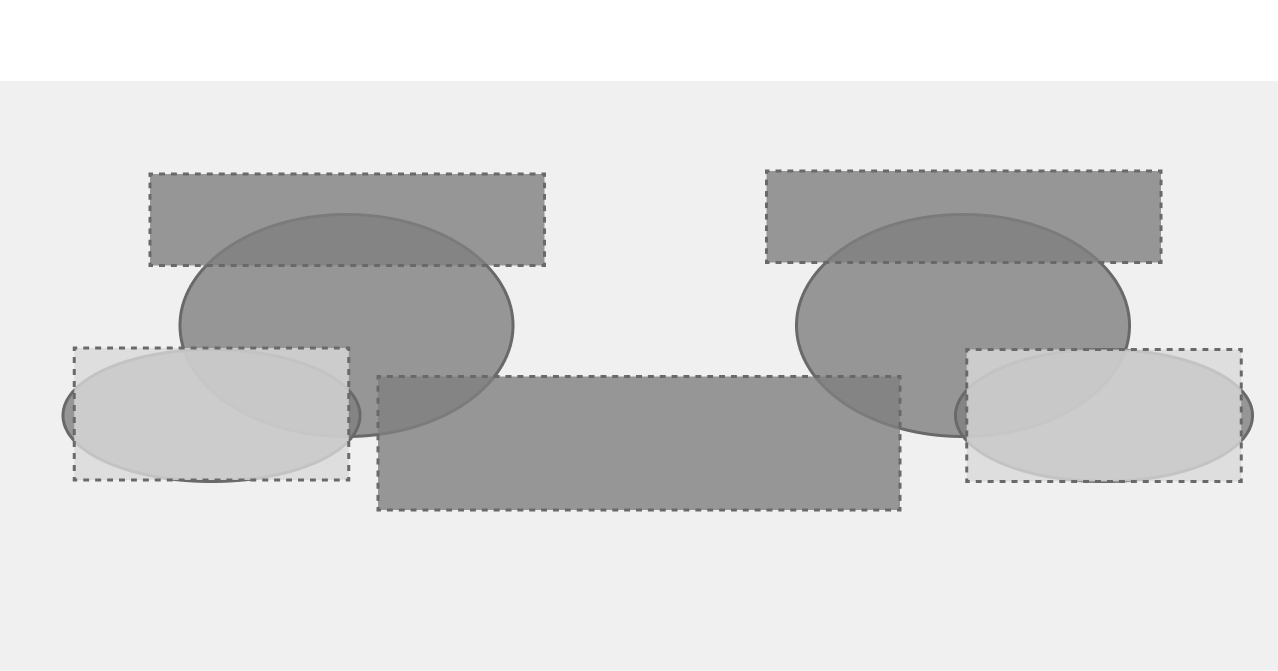}
    \\
    \midrule

    \multicolumn{2}{l}{\textbf{Personalized Face Generation Phase}} \\
    \leftpad
    1.\;@left-eyebrow-area @right-eyebrow-area thick middle, thin ends \lsep\newline
    2.\;@right-eye @left-eye simple black/white lines, medium sclera proportion \lsep\newline
    3.\;@left-blush @right-blush light pink, gradient dark center $\rightarrow$ light edge \lsep\newline
    4.\;@emoji-area blank before user interaction \lsep\newline
    5.\;@mouth-area simple black line
    &
\includegraphics[height=\imgH,keepaspectratio]{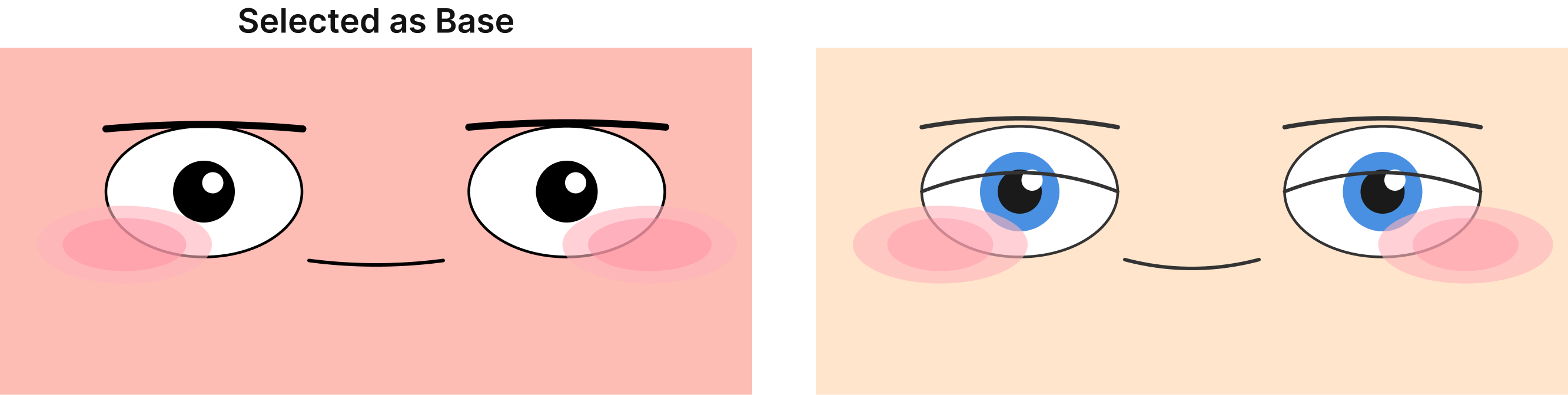}
    \\
    \midrule

    \multicolumn{2}{l}{\textbf{Contextual Facial Expression Generation Phase}} \\
    \leftpad
    1.\;@emoji-area generates emojis that change dynamically according to user emotion
    &
\includegraphics[height=\imgH,keepaspectratio]{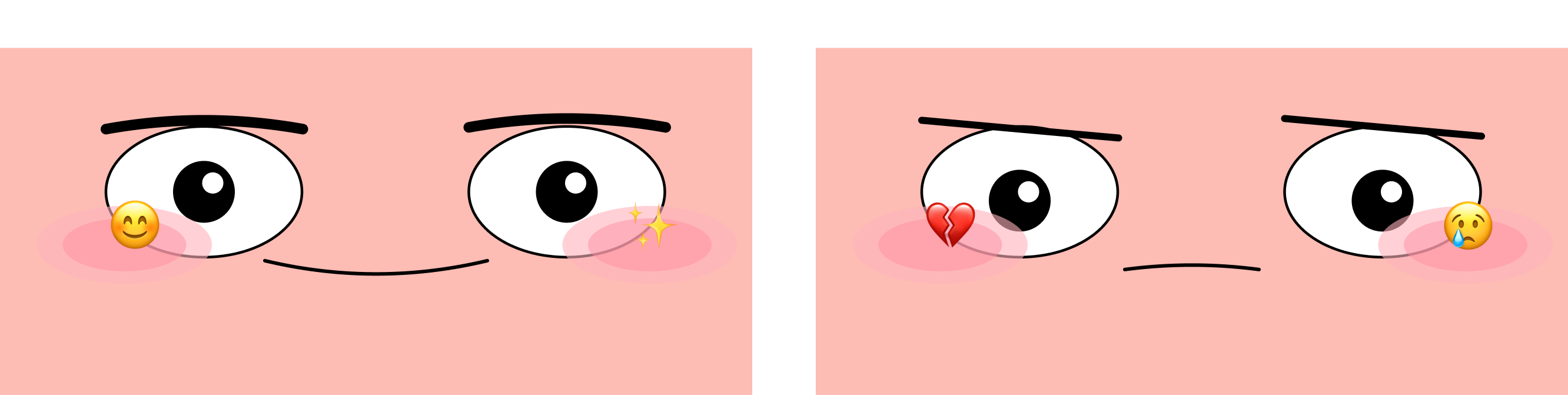}

    \\
    \bottomrule
  \end{tabular}

  \label{tab:task1-p1}
\end{table*}}{
\begin{table*}[t]
  \centering
  \caption{P1 generates characters with defined organs and context-related dynamic emojis in Task 1.}
  \footnotesize

  \setlength{\tabcolsep}{6pt}
  \renewcommand{\arraystretch}{1.12}

  \begin{tabular}[t]{p{0.58\textwidth} p{0.36\textwidth}}
    \toprule
    \textbf{Rules Definition} & \textbf{Output Visualization} \\
    \midrule

    \multicolumn{2}{l}{\textbf{Semantic Tags}} \\
    \leftpad
    1.\;Preset Tags: [face], [left-eye], [right-eye], [mouth-area], [emoji-area] \lsep\newline
    2.\;Custom Tags: [left-eyebrow-area], [right-eyebrow-area], [left-blush], [right-blush]
    &
    \topimg{assets/Task1-P1-1.png}
    \\
    \midrule

    \multicolumn{2}{l}{\textbf{Personalized Face Generation Phase}} \\
    \leftpad
    1.\;@left-eyebrow-area @right-eyebrow-area thick middle, thin ends \lsep\newline
    2.\;@right-eye @left-eye simple black/white lines, medium sclera proportion \lsep\newline
    3.\;@left-blush @right-blush light pink, gradient dark center $\rightarrow$ light edge \lsep\newline
    4.\;@emoji-area blank before user interaction \lsep\newline
    5.\;@mouth-area simple black line
    &
    \topimg{assets/Task1-P1-2.png}
    \\
    \midrule

    \multicolumn{2}{l}{\textbf{Contextual Facial Expression Generation Phase}} \\
    \leftpad
    1.\;@emoji-area generates emojis that change dynamically according to user emotion
    &
    \topimg{assets/Task1-P1-3.png}

    \\
    \bottomrule
  \end{tabular}

  \label{tab:task1-p1}
\end{table*}}

\renewcommand{\imgH}{2.5cm}

\begin{table*}[h]
  \centering
  \caption{P2 generates personalized cloud-shaped characters in Task 3.}
  \footnotesize
  \setlength{\tabcolsep}{4pt}
  \renewcommand{\arraystretch}{1.05}
  \begin{tabular}[t]{p{0.57\textwidth} p{0.37\textwidth}}
    \toprule
    \textbf{Rules Definition} & \textbf{Output Visualization} \\
    \midrule
    \multicolumn{2}{l}{\textbf{Semantic Tags}} \\
    1.\;Preset Tags: [left-eye], [right-eye], [mouth-area], [emoji-area] \newline
    2.\;Custom Tags: [body-area], [ribbon-area]
    &
    \vtop{\vskip0pt\hbox{\includegraphics[height=\imgH,keepaspectratio]{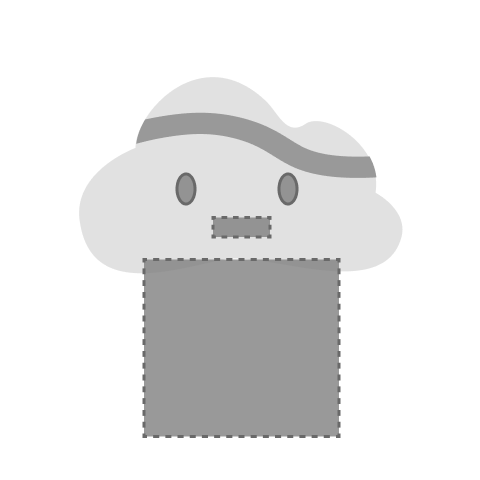}}}
    \\
    \midrule
    \multicolumn{2}{l}{\textbf{Personalized Face Generation Phase}} \\
    \textit{Design Rules:}\newline
    1.\;@emoji-area This is the carrier for the cloud character's emotions, where emojis can be randomly scattered within the area. \newline
    2.\;@mouth-area This is the region for mouth expression generation, where lines of different curvature can be combined to form expressions. \newline
    3.\;@ribbon-area This is the decorative part of the character, which can be filled with different colors. \newline
    4.\;@body-area This is the main body of the character, which can be filled with different colors (e.g., solid or gradient) and patterns (e.g., stripes or polka dots).
    \newline\newline
    \textit{Context Definition:}\newline
    Personality, Hobbies \newline\newline
    \textit{Context--Character Mapping Rules:}\newline
    1.\;@Personality The user's personality determines the body's color and pattern. For example, a quiet personality results in cool, low-saturation colors. \newline
    2.\;@Hobbies The user's hobbies determine the ribbon's color and pattern. For example, if the user likes basketball, the ribbon can be filled with a basketball texture.
    &
    \vtop{\vskip0pt\hbox{\includegraphics[height=\imgH,keepaspectratio]{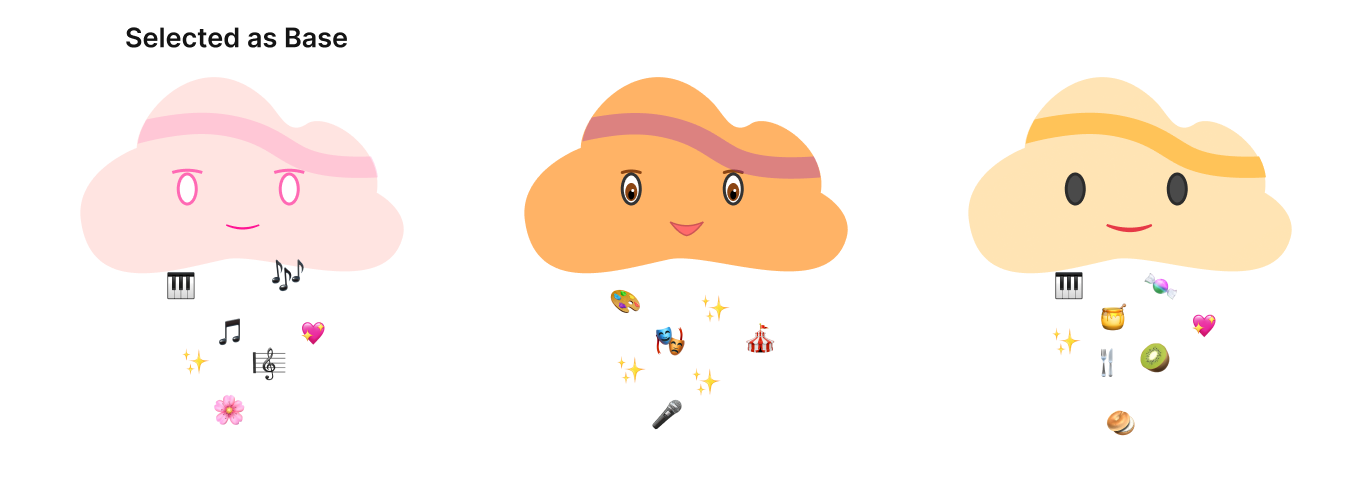}}}
    \\
    \midrule
    \multicolumn{2}{l}{\textbf{Contextual Facial Expression 
Generation Phase}} \\
    \textit{Design Rules:}\newline
    1.\;@left-eye @right-eye The eyes change their degree of openness according to the user's mood and behavior, e.g., half-closed when drowsy. \newline\newline
    \textit{Context Definition:}\newline
    Atmosphere, User Emotion \newline\newline
    \textit{Context--Expression Mapping Rules:}\newline
    1.\; The @Atmosphere controls the emoji's content and density. For example, if the user mentions autumn, falling leaves can be included. \newline
    2.\;@User Emotion @User Emotion changes the emoji's content and density. For example, if the user feels sad, many water droplets will appear in the area.
    &
    \vtop{\vskip0pt\hbox{\includegraphics[height=\imgH,keepaspectratio]{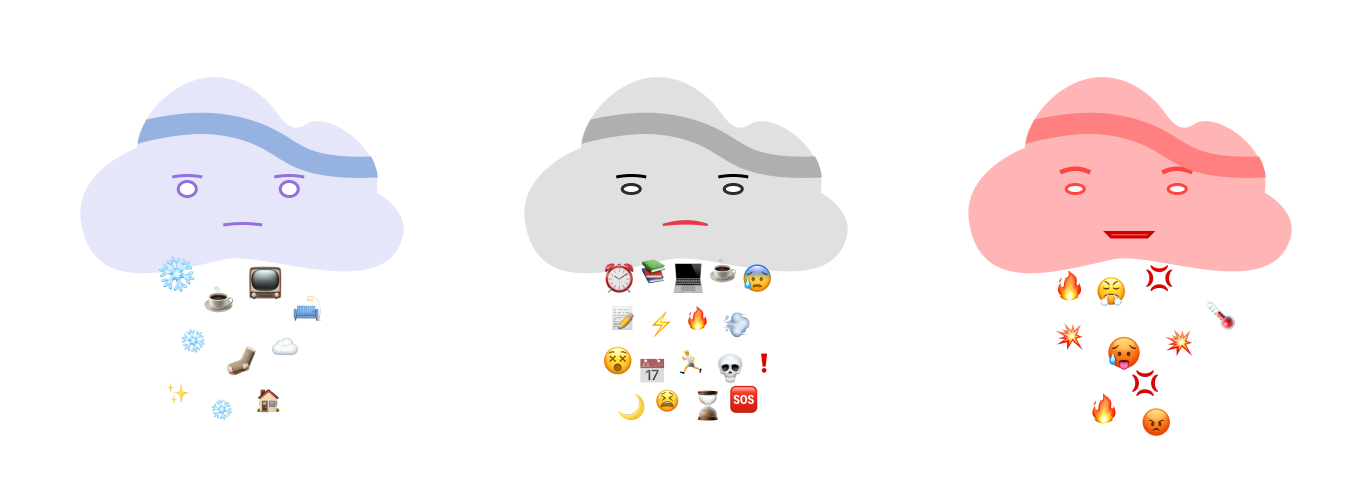}}}
    \\
    \bottomrule
  \end{tabular}
  \label{tab:task3-p2}
\end{table*}

\end{document}